\documentclass[12pt,preprint]{aastex}

\shorttitle{The low-z Carnegie Supernova Project}
\shortauthors{CSP et al.}

\begin{document}

\title{The Carnegie Supernova Project: The Low-Redshift Survey}
\author{Mario Hamuy\altaffilmark{1}, Gast\'on Folatelli, Nidia I. Morrell, Mark M. Phillips}
\affil{Carnegie Institution of Washington, Las Campanas Observatory, Colina El Pino s/n, Casilla 601, Chile}
\author{Nicholas B. Suntzeff} 
\affil{Cerro Tololo Inter-American Observatory, National Optical Astronomy Observatory\altaffilmark{2},
Casilla 603, La Serena, Chile} 
\author{S. E. Persson}
\affil{Carnegie Institution of Washington, 813 Santa Barbara Street, Pasadena, CA 91101, USA}
\author{Miguel Roth, Sergio Gonzalez, Wojtek Krzeminski, Carlos Contreras}
\affil{Carnegie Institution of Washington, Las Campanas Observatory, Colina El Pino s/n, Casilla 601, Chile}
\author{Wendy L. Freedman, D. C. Murphy, Barry F. Madore, P. Wyatt}
\affil{Carnegie Institution of Washington, 813 Santa Barbara Street, Pasadena, CA 91101, USA}
\author{Jos\'{e} Maza}
\affil{Departamento de Astronom\'\i a, Universidad de Chile, Casilla 36-D, Santiago, Chile}
\author{Alexei V. Filippenko, Weidong Li}
\affil{Department of Astronomy, University of California, 601 Campbell Hall 3411, Berkeley, CA 94720-3411}

\and

\author{P. A. Pinto}
\affil{Steward Observatory, The University of Arizona, Tucson, AZ 85721}

\altaffiltext{1}{Current address and E-mail: Departamento de Astronom\'\i a, Universidad de Chile, Casilla 36-D, Santiago, Chile (mhamuy@das.uchile.cl)} 
\altaffiltext{2}{Cerro Tololo
Inter-American Observatory, National Optical Astronomy Observatory,
operated by the Association of Universities for Research in Astronomy,
Inc., (AURA), under cooperative agreement with the National Science
Foundation.}

\begin{abstract}

Supernovae are essential to understanding the chemical evolution of the 
Universe.  Type~\,Ia supernovae also provide the most powerful observational 
tool currently available for studying the expansion history of the Universe and the
nature of dark energy.  Our basic knowledge of supernovae comes from the 
study of their photometric and spectroscopic properties.  However, the 
presently available data sets of optical and near-infrared light curves of 
supernovae are rather small and/or heterogeneous, and employ photometric 
systems that are poorly characterized.  Similarly, there are relatively 
few supernovae whose spectral evolution has been well sampled, both in 
wavelength and phase, with precise spectrophotometric observations.  The 
low-redshift portion of the Carnegie Supernova Project (CSP) seeks to remedy
this situation by providing photometry and spectrophotometry of a large
sample of supernovae taken on telescope/filter/detector systems that are
well understood and well characterized. During a five-year program which began in September
2004, we expect to obtain high-precision $u'g'r'i'BVYJHK_s$ light curves
and optical spectrophotometry for about 250 supernovae of all types.
In this paper we provide a detailed description of the CSP survey observing
and data reduction methodology.  In addition, we present preliminary
photometry and spectra obtained for a few representative supernovae during
the first observing campaign.

\end{abstract}

\keywords{supernovae, cosmology }

\section{INTRODUCTION}

A universe dominated by normal mass should undergo deceleration as it
expands.  Thus, the counter-intuitive discovery of an accelerating universe 
based on observations of Type~\,Ia supernovae \citep{riess98,perlmutter99} 
is of profound significance for physics.  Evidently ``dark energy,'' in the
form of Einstein's cosmological constant or a more general scalar energy 
field, is the dominant mass/energy constituent of the Universe today.

These important implications depend critically on the quality of the light 
curves of the Type~\,Ia supernovae (SNe~Ia, hereafter) and the ability to 
K-correct, deredden, and normalize these to a standard luminosity.  The 
evidence for an accelerating universe is based on a differential measurement 
between local and distant SNe~Ia (at lookback times of 4--10 Gyr).
The local samples are very 
heterogeneous, and as more SNe have been added, the full sample dispersion 
around the local Hubble flow has increased from 0.12 \citep{phillips99} to 
0.18 mag \citep{jha02} in units of distance modulus.  Moreover, there are
still legitimate concerns about possible systematic errors due to 
poorly understood photometric systems \citep{suntzeff00,stritzinger02}, 
reddening corrections \citep{phillips99}, and evolutionary effects 
\citep{hamuy00,gallagher05}.  A new and larger sample of nearby 
($z < 0.07$) SNe, where these sources of observational
error have been duly accounted for, is urgently needed.

With that purpose in mind, we have initiated a five-year
program, the Carnegie Supernova Project (hereafter CSP), to obtain
well-calibrated optical and near-infrared light curves as well as 
optical spectrophotometry of $\sim$250 Type~\,Ia and core-collapse SNe.
The CSP is built upon the unique resources of the
Las Campanas Observatory (LCO) in Chile. We have guaranteed access to
large numbers of nights on the Swope 1~m and the
duPont 2.5~m telescopes ($\sim$300 per year in both telescopes together),
which are equipped with high-performance CCD 
optical imagers, near-infrared (NIR) cameras, and CCD optical 
spectrographs. In addition to providing densely sampled 
light curves covering the near-ultraviolet 
to the NIR ($u'g'r'i'BVYJHK_s$), we have the
means to obtain optical spectrophotometry at approximately weekly
intervals. The CSP is a follow-up project and relies on SNe discovered in
the course of other surveys. A large fraction of our targets come from the Lick 
Observatory Supernova Search \citep{li00,filippenko01,filippenko03,filippenko05}
conducted with the Katzman Automatic Imaging Telescope (KAIT), and from dedicated
SN searches by amateur astronomers (e.g., Tim Puckett, Tom Boles, Berto Monard,
Koichi Itagaki), which constitute a growing 
source of nearby SNe. The targets selected for the follow-up observations 
by the CSP are SNe discovered before or near maximum light with
$z \lesssim 0.07$ and $\delta$ $\lesssim$ +20$^\circ$.

The primary goal of the CSP is to establish a fundamental data set of 
optical and NIR light curves in a well-defined and well-understood photometric system
for all types of SNe.  A secondary goal is to obtain
complementary optical spectrophotometry for these same SNe. 
The data set for the Type~\,Ia events will allow us
to improve extinction corrections and to investigate systematic effects possibly due
to differences in age and metallicity. The data for the Type~\,II\,~SNe will
be used to establish and refine precise techniques for measuring
luminosity distances employing the Expanding Photosphere Method
\citep{eastman96, schmidt94, hamuy01, leonard02, dessart05} or the Standardized Candle
Method \citep{hamuy02}, thereby providing an independent check on the
Type~\,Ia results.  We will be able to explore the use of both SN types
for studies of local galaxy flows and independently measure the
convergence depth (the distance at which the bulk flows smooth out
into the so-called large-scale Hubble flow). 
Ultimately, the data set will serve as a reference for observations of
distant SNe that will be obtained in coming years in the course of the 
Joint Dark Energy Mission \citep{aldering05} and those
being obtained in the course of the high-$z$ SN
surveys such as the CFHT Legacy project \citep{pritchet05}, ESSENCE
\citep{matheson05}, and our own high-$z$ component of the CSP which seeks to
measure rest-band $I$ magnitudes of SNe~Ia at $z \approx 0.5$ using the
Magellan telescopes \citep{freedman05}.

The low-$z$ CSP data set will also allow us to gain a deeper understanding of the
physics of thermonuclear (Type~\,Ia) events and the different classes of 
core-collapse SNe (Types~\,II,~\,Ib,\,~Ic). For example, during the first CSP observing
campaign, we obtained excellent coverage of SN\,~2005bf, a peculiar, luminous
Type~\,Ic event which peaked 35 days after explosion and which may
represent a 
transition object between the SNe associated with gamma-ray bursts and 
ordinary SNe~Ib \citep{folatelli05}.

The purpose of this paper is to describe the low-$z$ CSP experiment,
to explain the general procedures for data acquisition and reduction,
to summarize the results obtained during the first (2004--2005) low-$z$ CSP
observing campaign, and to present the data for a few representative SNe.
In $\S$ \ref{inst} we discuss the instrumentation, in $\S$
\ref{obs} we describe our observations, in $\S$ \ref{red} we
explain the data reduction procedures, and in $\S$ \ref{res} we
show representative light curves and spectra obtained thus far.

\section{TELESCOPES, INSTRUMENTS, AND PHOTOMETRIC SYSTEMS}
\label{inst}

A large fraction of our targets come from the Lick Observatory
Supernova Search (LOSS) which discovers approximately 50\%
of the low-$z$\,~SNe found annually.  Although LOSS is conducted at
northern latitudes, in 2004 the search was adjusted to include more
galaxies south of the celestial equator, and thus accessible to the Carnegie
telescopes.  Over half of the current LOSS SNe are located south of
$\delta$=+20$^\circ$ and are observable from LCO. The discovery redshift 
limit, $z \lesssim 0.05$, of the LOSS SNe is ideally matched to the telescopes
and detectors available to the CSP. The rest of the targets come from
a variety of searches carried out by amateur astronomers at different 
observatories. The telescopes employed to date for the CSP are the Swope 
1.0~m, duPont
2.5~m and, to a lesser extent, the Magellan 6.5~m telescopes at LCO. 
In addition, some spectra have been obtained with 
the 1.5~m telescope at Cerro Tololo Inter-American Observatory (CTIO). 
Various telescopes at Lick Observatory have also been used for photometry
and spectroscopy (e.g., Foley et al. 2003), but those
observations will be described in separate papers.

In the remainder of this section, we describe the instrumentation used
with these telescopes.
Table \ref{tab_inst} lists the instruments used for the low-$z$ CSP.

\subsection{Optical Imaging}
\label{op_img}

The vast majority of the CSP optical imaging is being obtained with the 
Swope 1~m $f$/7 telescope Direct CCD Camera, which uses a 
$2048 \times 3150$ pixel, 15 $\mu$m pixel$^{-1}$ SITe CCD.
This detector has a readout noise of 6.6 e$^-$ and a gain of 
2.5 e$^-$~ADU$^{-1}$ (where ADU means analog-to-digital units).  
To speed up the CCD readout 
and save disk space, we limit the readout to $1200 \times 1200$ pixels. 
At a scale of $0.435''$ pixel$^{-1}$, this corresponds to a field of view 
(FOV) of $8.7' \times 8.7'$.  The quality of images 
obtained at the Swope telescope ranges from $\sim$1--$2''$ full 
width at half-maximum (FWHM), with an average seeing of $1.3''$.

For our optical passbands we utilize Sloan Digital Sky Survey (SDSS)
$u',g',r',i'$ filters in addition to Johnson $B$ and $V$ filters
\citep{fukugita96,bessell90}. We have chosen SDSS filters because we feel it
is likely that this will be the dominant photometric filter set for the
next decade due to the all-sky photometric maps of SDSS (north) and the
Mount Stromlo and Siding Spring Observatories SkyMapper\footnote{http://msowww.anu.edu.au/skymapper}
project (south). We decided not to use the $z'$ filter because the red side of the filter is
determined by the CCD quantum efficiency (QE) and not solely by the filter bandpass.
The CCD red response is very temperature sensitive, 
causing the combined filter-plus-detector
throughput to be variable.  The $z'$ bandpass also overlaps with a saturated
H$_{2}$O band at 9,300 \AA, adding further uncertainty to its total throughput and
effective wavelength. The $B$ and $V$ filters were included in
our program to sample the spectral region covered by the $g'$ filter with 
somewhat narrower filters, and to facilitate comparison of our results with 
historical data sets.

\begin{figure}
\plotone{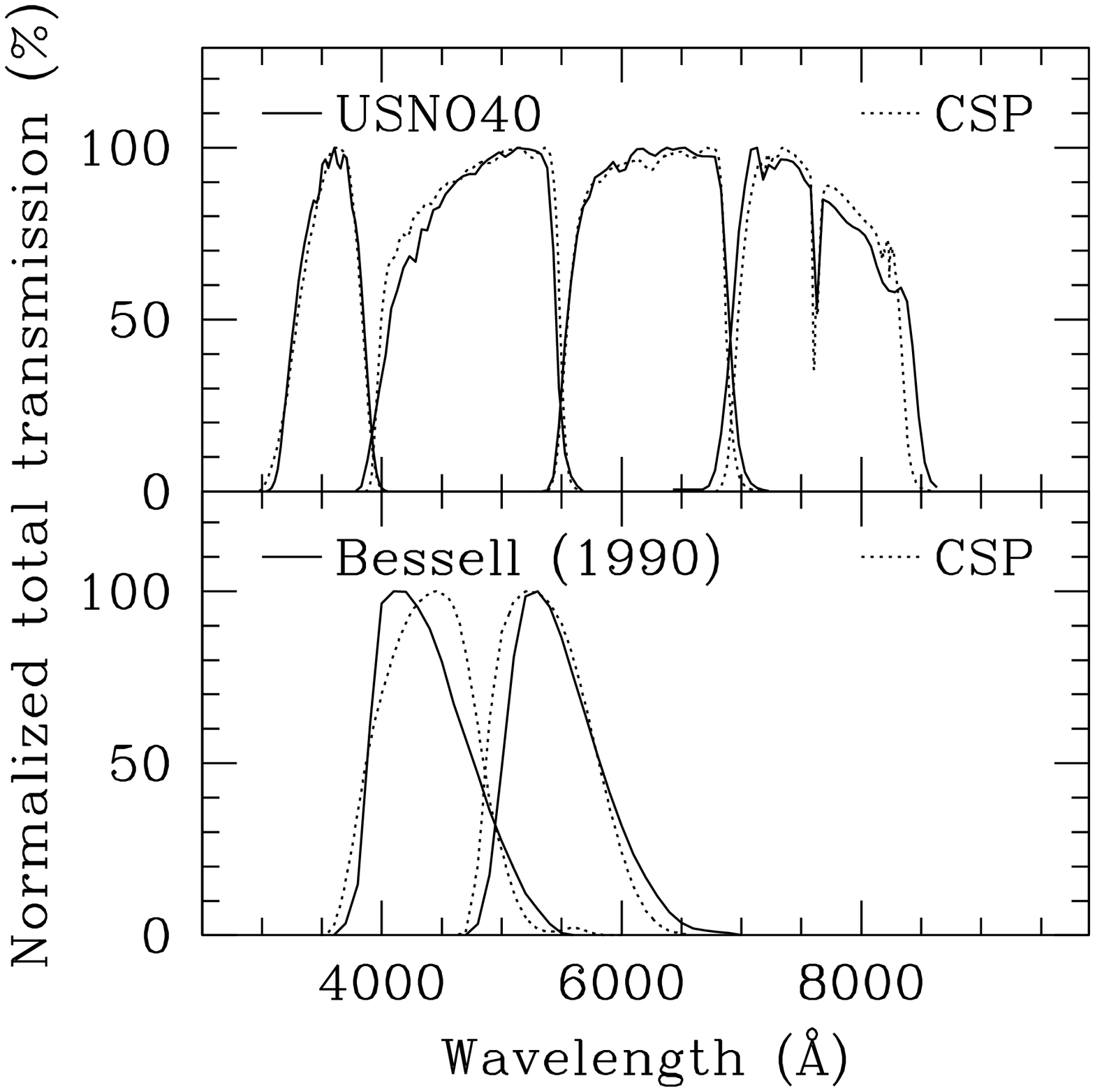}
\caption{(upper) Natural system synthetic bandpasses for the Swope $u',g',r',i'$ SDSS filters
(dotted lines) and the USNO 40-inch telescope standard bandpasses (solid lines) used for the establishment
of the SDSS photometric system (normalized to 100\% at maximum).
(bottom) Natural system bandpasses for the Swope Harris $B$ and $V$
filters (dotted lines) and the $B$ and $V$ Bessell (1990) bandpasses
(divided by a linear function in wavelength and normalized to 100\% at maximum).
\label{filters_fig}}
\end{figure}

Our SDSS filters were manufactured by Asahi Spectra Company 
Limited\footnote{This is the same company that built the filters used with
the US Naval Observatory (USNO) 40-inch telescope for the establishment of the
SDSS photometric system \citep{smith02}.} as specified in Table \ref{tab_filters}.
We have synthesized natural system passbands by combining the filter transmissions
with the CCD QE, two aluminum reflectivity curves (one for the primary and another for
the secondary mirror), and an atmospheric transmission spectrum.
The resulting SDSS bandpasses are shown in the upper panel of Figure 
\ref{filters_fig} along with the standard bandpasses (normalized at 
maximum) for the USNO 40-inch telescope.
This comparison reveals an excellent match between both systems,
with the exception that our $i'$ bandpass is somewhat narrower than that used
at the USNO; the effective wavelength is, however, unchanged.
The bottom panel of Figure \ref{filters_fig} compares the passbands for
our Harris $B$ and $V$ filters with the Johnson $B$ and $V$ passbands
described by \citet{bessell90}. It is clear that our instrumental system provides
a reasonable match to the Johnson system. 

Since one of our goals is to eliminate systematic errors in the SN
magnitudes caused by differences between the instrumental and the
standard bandpasses, we have started a program of regular measurements 
of the transmission of our filters using a spectrometer at LCO.
Similarly, we are planning in the near future to improve the measurement of
the QE curve for the Swope CCD using equipment available for this purpose 
at the CTIO laboratories. Our natural-system
bandpasses will be regularly updated as we improve our measurements
of the CCD QE, the mirror reflectivities, and other optical elements
of our instrument. All of these will be posted on the CSP web site
and made available through the NASA/IPAC Extragalactic Database (NED).

Some $BVI$ imaging has also been obtained with the Wide Field 
Re-imaging CCD Camera (WFCCD) on the 2.5~m $f$/7.5 duPont telescope. This
uses a $2048 \times 2048$ pixel Tektronix CCD with 
24 $\mu$m pixels.  A subraster is used to observe a $23' \times 13'$ 
FOV at a scale of $0.774''$ pixel$^{-1}$. 

Likewise, a small amount of $BVR$ imaging has been taken with the 
Low-Dispersion Survey Spectrograph (LDSS-2) \citep{allington94} on the 
6.5~m $f$/11 Magellan Clay telescope.  Until the end of 2004, this
instrument employed a SITe CCD for which we limited the readout to
a section of $2048 \times 1240$ pixels containing a circular FOV
of $6.4'$ diameter. The pixel size of 15 $\mu$m corresponds to a
scale $0.38''$ pixel$^{-1}$. The $BVR$ filter specifications for this
instrument are given in Table \ref{tab_filters}. In February 2005, 
the LDSS-2 instrument
underwent a significant upgrade which involved replacement of both the
optics and the detector.  The new CCD is an STA0500A detector with 
$4064 \times 4064$
pixels which, in combination with the new optics, has a scale of
$0.19''$ pixel$^{-1}$.  The upgraded instrument, now called LDSS-3,
 retains the same $B$ filter used in
LDSS-2, but the $V$ and $R$ filters were replaced by SDSS $g',r',i'$ and $z'$
filters.  Although we have not yet employed LDSS-3 for the CSP, we expect
to do so occasionally in the future.

For galaxy subtraction, we need template images after the SN has
faded. These will be obtained beginning in November 2005 
with the Direct CCD Camera on the 2.5~m duPont telescope.  
The detector is the same one used with the WFCCD, 
and covers a FOV of $8.8' \times 8.8'$ with
a scale of $0.259''$ pixel$^{-1}$.  With a typical image quality of 
$0.7''$ (FWHM) at the duPont telescope, 
this instrument is ideal for acquiring 
high-quality deep templates. The filter set will be the same as that
used on the 1~m telescope.

\subsection{Infrared Imaging}

During the first CSP campaign, we obtained 85\% of the NIR images
with the Wide Field Infrared Camera (WIRC) on the 2.5~m duPont
telescope \citep{persson02}, but this situation will change when a new NIR
camera, RetroCam, becomes available at the Swope telescope for the second
campaign (see $\S$ \ref{future_sec}). WIRC is equipped with four $1024 
\times 1024$ pixel Rockwell
HAWAII-1 HgCdTe arrays forming a $2 \times 2$ square footprint 
with a $175''$ center-to-center spacing in the reimaged telescope focal plane.
Each array covers a FOV of $\sim3.3' \times 3.3'$ with a
$0.196''$ pixel$^{-1}$ scale. For SN imaging we use the two best 
quality detectors of the four (Detectors \#2 and \#3).
Departures from linearity in measured flux are 1\% below 30,000 e$^-$,
growing to about the 5\% level at $\sim$60,000 e$^-$
(see $\S$ \ref{nir_red}). The typical image quality for this 
instrument falls between $0.5''$ and $0.8''$ (FWHM).
We observe in the $Y$, $J$, $H$, and $K_s$ passbands, using filters manufactured
by Barr Associates Ltd. ($Y$, $J$, $H$), and Optical Coating Laboratory Inc. 
($K_s$).

Some NIR data has also been obtained using the Persson Auxiliary Nasmyth
Infrared Camera (PANIC) mounted on the 6.5~m Baade telescope \citep{martini04}.
This instrument has a single $1024 \times 1024$ pixel 
HgCdTe HAWAII-1 detector covering
a $2' \times 2'$ FOV at a scale of $0.125''$ pixel$^{-1}$.
The non-linear behavior of the detector is similar to that
of the WIRC arrays. The active optics on the primary mirror produce a typical
image quality of 0.3--$0.6''$ (FWHM). The $Y$, $J$, $H$, and 
$K_s$ filters are similar to those employed with WIRC.

Figure \ref{IR_filters_fig} shows the total system transmission curves
for WIRC and PANIC, with the $YJHK_s$ filters. 
The plotted curves include atmospheric transmission, throughput of every optical element in the
telescope and instrument, and detector QE. The graph
shows excellent agreement between the WIRC and PANIC systems.

\subsection{Spectroscopy}

The majority of our spectroscopic observations have been obtained with the 
2.5~m duPont telescope using the WFCCD instrument in its spectroscopic 
long-slit mode.  A 400 line mm$^{-1}$ blue grism is employed with the Tektronix
$2048 \times 2048$ pixel CCD 
to provide a wavelength coverage from 3,800--9,200~\AA\ at 
a dispersion of 3.0~\AA~pixel$^{-1}$.  For the 
$\sim1.6''$ slit width used for the SN observations, this setup gives 
a FWHM resolution of $\sim$6.0~\AA.  For a $V$=16 magnitude object in clear 
conditions, a signal-to-noise ratio (S/N) per pixel of 50 at 
5,000~\AA~is typically achieved for an exposure time of 1,200~s. 

When the WFCCD is not available on the 2.5~m telescope due to block 
scheduling constraints, it has been possible to obtain some spectra with
the Las Campanas Modular Spectrograph.  This instrument uses a SITe
$1752 \times 572$ pixel CCD with 15 $\mu$m pixel$^{-1}$ and a 
300 line\,mm$^{-1}$ grating (blazed at 5,000~\AA). The resulting spectral 
coverage is $\sim$3,800--7,300~\AA~at a dispersion of 
2.45~\AA~pixel$^{-1}$.  A slit width of $1''$ is used for the SN 
observations, which gives a FWHM resolution of $\sim$7~\AA.  Compared to
the WFCCD, this instrument has lower efficiency and longer exposures are 
required.

Occasionally during the first CSP campaign, we were scheduled single nights at 
the 6.5~m Magellan Clay telescope with LDSS-2.  For these observations,
a 300 line mm$^{-1}$ grism blazed at 5,000~\AA~was employed, providing 
wavelength coverage of 3,600--9,000~\AA\ at a dispersion of 
5.3 \AA~pixel$^{-1}$.  For a $1''$ slit, this translates to a FWHM
resolution of $\sim13.5$~\AA.

Approximately 10\% of the CSP spectroscopy has been carried out with the 
CTIO 1.5~m telescope in the service mode of the SMARTS 
consortium\footnote{http://www.astro.yale.edu/smarts/}.  We use the facility 
Ritchey-Chr{\'e}tien Cassegrain Spectrograph 
equipped with a $1200 \times 800$ pixel LORAL CCD,
usually at $R=300$ which gives a dispersion of 5.7~\AA~pixel$^{-1}$,
a wavelength coverage of 3,000--10,100~\AA, and a FWHM resolution of 
$\sim$14~\AA. 

Note that for none of the above spectrographs do we employ a filter to block 
second-order light, which means that
there could be second-order contamination redward of about 7,000~\AA.
This effect will be quantified in a future paper.

\section{OBSERVATIONS}
\label{obs}

Our plan is to carry out an intensive nine-month campaign each year, September
through May, for five years. The extended nature of each campaign
allows us to follow SNe discovered before the end of month seven (March).
Given that various searches are currently producing $\sim$100 
southern low-$z$ SNe per
year, we expect complete light-curve coverage for $\sim$50 SNe (25 SNe~Ia,
20 SNe~II, and 5 SNe~Ibc) for each nine-month campaign.
Our observational goals are: (1) for SNe~Ia
and SNe~Ibc, to acquire $u'g'r'i'BVYJHK_s$ photometry to 0.03 mag
precision (random error) every 3--4 days, and optical spectroscopy
every 5--7 days from discovery through 50 days past maximum; (2) for
SNe~II, the same, except for additional coverage through the $\sim$100-day
extended plateau phase and onset of the nebular phase, but with a
cadence of 7--10 days.

We collected data with no major interruption during our first campaign between
September 2004 and May 2005. During this period, the CSP was allocated
190 nights on the 1~m telescope for optical imaging. We also had 57 nights
on the 2.5~m telescope, of which 33 were for NIR imaging, 19 for optical 
spectroscopy, and 5 for optical imaging.  Strict observing procedures
were established in order to obtain the data in the most homogeneous manner 
possible.

As soon as a report of a SN candidate was received, we triggered the 
follow-up program to begin generally the same night, even without a 
spectroscopic type or
age. If a spectrum later revealed that the SN had been caught
after maximum light we dropped the object from our list of
targets. Sometimes it took a while ($\lesssim$2 weeks) to obtain the
SN spectrum, and in such cases, we used the photometric information collected
by us to evaluate its probable type or age. 
For these purposes we devised a pipeline
to produce preliminary light curves. At any given time we generally had 15
objects on our target list.

We have divided the labor of our project into three working groups,
each one having two individuals responsible for
(1) maintaining the observing instructions and procedures up to date, 
(2) sending the prioritized list of targets to the observers, 
(3) developing data reduction software, and 
(4) reducing the data immediately following the night.
Observers were responsible for
(1) checking the observing program for the night,
(2) taking data and calibration images,
(3) saving the data on storage media,
(4) preparing a nightly observing report, and
(5) updating the observing log for each SN.
A fluid interaction and communication between the working group leader
and the observers allowed us to optimize the telescope time, detect
problems with the instrumentation, obtain excellent data, and reduce
them promptly.

We designed and implemented a web site for the project which can
be found at \\
http://csp1.lco.cl/$\sim$cspuser1/. The SN list,
finding charts, observing procedures, observing programs, data reduction
manuals, and many useful tools are available at this URL. This web
site is also open to the public, where the
photometric evolution of the objects we are following can be seen.

In what follows we describe the specific observations that each of the
working groups gathered during the first low-$z$ CSP campaign.

\subsection{Optical Imaging}

Optical imaging with the 1~m Swope telescope would typically begin
during the daytime by taking bias frames, as well as
dome flats for each of the $g'r'i'BV$ filters with exposure
times chosen to achieve $\sim$23,000 e$^-$ per pixel.  Immediately
after sunset, we would observe the twilight sky with the $u'$ filter, followed by
sky flats with two or three additional filters. Generally, 3--5 images were 
obtained per filter, making sure to offset the telescope between the individual
images. The telescope was then focused using the
$V$ filter. For the remaining five filters, we would use previously derived
focus offsets relative to the $V$ filter.

During the night, SNe were observed according to the priorities previously
assigned by the head of the optical-imaging working group, who
determined exposure times based on the source brightness (typically 
$15 \lesssim V \lesssim 20$ mag) and our requirement to achieve 
3\% photometry.  In general we would observe the SNe in all six filters, taking 
one image per filter.  However, given that we usually had more objects to 
observe
than telescope time available, we would drop the 3\% precision requirement
for the faintest objects and, in such cases, limit the exposure time to 900 s.
When the resulting photometric precision reached 0.2 mag, we discontinued the
observations through that filter. This would happen first in the $u'$ filter,
typically at a magnitude of $\sim$21.

Johnson and SDSS standards \citep{landolt92,smith02} were observed regularly 
during photometric nights. At the beginning of the first campaign, few science
targets were available so we observed 11--13 standard fields per night. As the
campaign unfolded and more targets became active, we limited these observations
to 4--5 fields. In all cases, we spread the observations of the standards over 
an airmass range of 1--2 for extinction determinations. The exposure times 
were 3--120~s depending on the filter used and the brightnesses of 
the stars.  The purpose of these observations was to calibrate local
standard stars in the field of each SN so that later on we could do differential
photometry of the SN relative to them.

During cloudy nights, we performed detailed shutter timing and 
linearity tests on the Swope telescope. The results can be found in
Appendix \ref{shutter_app} and Appendix \ref{linearity_app}.

The observations with WFCCD and LDSS-2 were performed in a way similar to those
with the Swope telescope, except that we did not observe standard stars. 
We expect to observe standard fields in the future with the purpose of
deriving 
color terms for the filters used. We did not measure the CCD responses and
the shutter corrections for these instruments, but we expect to do so in the
future for the WFCCD (the LDSS-2 is no longer operational).

\subsection{Infrared Imaging}

Calibration images for WIRC and PANIC were usually taken just after
closing the dome at the end of the night.
Typically the observer would obtain 
15 dark frames with exposure times
matching those of the science objects, 15 dome flats per filter with
the dome lights on  (with $\sim$20,000 e$^-$~pixel$^{-1}$), followed by
15 dome flats with the lights off. Final dome flats were constructed from 
the ``lights-on'' images minus the ``lights-off'' images. 
This procedure was found to be
preferable, especially at $K_s$, to using dark-subtracted ``lights-on'' dome
flats since contributions
to the flux from additive sources were removed correctly. The latter sources
include the camera window, telescope mirrors, and scattered thermal emission
which illuminate the detector differently than does the flat-field screen.

For WIRC, we obtained twilight flats on some
nights to test the illumination quality of the dome flats. These
tests showed agreement between twilight and dome flats within 1\% for
all four NIR passbands. However for PANIC, it was necessary to take
twilight flats for illumination corrections to bring the photometric
flatness to better than 2\%. We took twilight flats every night in
sets of at least five exposures per filter, offsetting the telescope
between exposures, and controlling the flux level to be in the linear
regime. We used dome flats for both WIRC and PANIC to build bad-pixel
masks for each night.

For the WIRC observations, we preselected 
the telescope pointing for each SN such that
there were a number of good local 
calibration stars in the field. Thus the SN was not
always centered in the detector. We started observing each SN by placing it on 
Detector \#2 and taking a sequence of dithered exposures in one of the filters.
Usually we used nine dither positions in a $3 \times 3$ square pattern, but when
the SN was bright enough we used only five positions. We set the exposure time
of the individual images to 20, 30, or 45 s, depending on the
brightness of the SN. Occasionally, for the faintest SNe, we 
repeated the exposures at each dither position. Once the dither sequence was
completed, we would offset the SN field to Detector \#3 and repeat the 
sequence for the same filter. This procedure allowed us to obtain sky images
suitable for subtraction for each of the two detectors while observing
the object on the other detector.  We usually observed with
all four filters, though in the case of faint SNe, we dropped the $K_s$
filter because of the large uncertainties introduced by the high
background levels that characterize this band. 
Typical times spent on each target, including overhead,
were between one and two hours, with effective exposures between 200 and 800 s 
in each filter.

There are advantages and weaknesses to this technique. The advantage
is that the nearby galaxies tend to be very large, making it generally
impossible to use the median of the dithered SN+galaxy data on a single
detector to form a clean sky image with no print-through. The sequence
of off-source dithered exposures (obtained when the SN+galaxy move to the other detector),
on the other hand, produces a very clean sky image. The disadvantage is that sometimes the 
sky may be taken as much as 15 minutes before or after the SN+galaxy exposure, and the NIR 
sky can change over this time. Thus the sky images, while very clean, may have 
larger residual subtraction features because of changing sky emission levels.

On the photometric WIRC nights, we observed three to five
standard stars from \citet{persson98} on Detector \#2.
For observing efficiency, we based the
photometric transformations to the standard system on data obtained with this
single detector. Standard stars were observed with five dither positions and three
repetitions at each position. We chose stars with $J$ fainter than mag
11 to avoid saturation and fixed the exposure time of individual
images to 5 s, resulting in total exposure times of 75 s for each
filter. The photometric transformations derived from these
observations were applied to comparison stars in the SN fields (observed
only with Detector\#2) to obtain precise magnitudes for these local standards.

For our PANIC observations, we selected a slightly different pointing for
each SN than that used for WIRC in order to include the maximum 
number of comparison stars in the smaller FOV of PANIC.
Observations were made with five dither positions and two or three exposures
per position, depending on the SN brightness. The exposure times
for individual images were 10, 20, or 30 s with total exposure times
ranging between 100 and 450 s. We followed each of these sequences 
by a sky sequence after offsetting the telescope by one or two arcmin
in any direction, making sure that the host galaxy fell outside of the 
sky field. 

\subsection{Spectroscopy}

Our spectroscopic observations typically started during the afternoon by taking
bias frames, followed by a series of dispersed dome flats with wide 
($\sim$7$''$)
and narrow ($\sim$1$''$) slits. In addition, for the WFCCD and
LDSS-2 instruments, dome flats were obtained for the respective imaging
filters ($BVI$ for the WFCCD, and $BVR$ for LDSS-2). 
We adjusted the lamp
voltage and the exposure times to obtain 10,000 e$^-$~pixel$^{-1}$ in
our flat-field direct images, and a maximum of 15,000 e$^-$~pixel$^{-1}$
in the spectroscopic flats.

We observed SNe with the narrow slit aligned along the parallactic angle
\citep{filippenko82} according to a priority list built for every night.
Total exposure times with the WFCCD and the Modular Spectrograph
varied between 900 and 2700 s, and between 180 and 900 s with LDSS-2. For each
SN, we divided the total exposure into three independent integrations for 
cosmic-ray rejection. In between the first two or last two exposures, 
an image of a comparison lamp for wavelength calibration was taken. During the night 
we observed at least two
spectrophotometric standards \citep{hamuy92,hamuy94b} with the wide slit. 
A few weeks after the beginning of the first campaign, we decided
to include one observation per night of a telluric standard \citep{bessell99}
with high S/N and the same narrow slit used for the observations of the 
SNe. Since data from the CTIO 1.5~m telescope were obtained
in service mode, we were not able to obtain such a calibration with that
instrument.

\section{DATA REDUCTION}
\label{red}

\subsection{Optical Imaging}

To accelerate and automate the reduction of the optical imaging, 
we developed a custom IRAF\footnote{IRAF is distributed by the National 
Optical Astronomy Observatories,
which are operated by the Association of Universities for Research
in Astronomy, Inc., under cooperative agreement with the National
Science Foundation.} script package to handle the Swope 1~m data.
For all nights, we first corrected the exposure time for the shutter timing
error (see Appendix \ref{shutter_app}).
Next we processed the images
through bias subtraction, non-linearity corrections (see 
Appendix \ref{linearity_app}),
and flat-field division. For a given filter, we constructed the flat field by
dividing the median-filtered combined sky flat by the median-filtered combined dome
flat, heavily smoothing this ratio, and multiplying this illumination
correction into the dome flat. This procedure removed a small
gradient of $\sim$1\%. When a nightly sky flat could not be obtained we
used one from a previous night.  Random checks of science images taken during
the night typically showed that the sky level across the images varied
by less than 1\%. Although fringing is present in the $i'$ band we did
not attempt to remove it from our frames. Since our 
$i'$-band instrumental magnitudes of the local standards in the SN fields can be
transformed to the standard system (as explained below) to better than 0.015 mag, this implies
that neglecting the fringing correction introduces an error $\lesssim$ 0.015 mag.

Next we used the IRAF {\em daophot} package to compute instrumental magnitudes 
for the standard stars with an aperture of $7''$ in radius (the same used in the
establishment of the standard system), and a sky annulus located 7--$9''$
from the star. We computed instrumental errors using a Poisson model based on
the noise parameters of the CCD. This model provided realistic errors for faint
stars, but excessively small errors
for bright stars. We adopted 0.015 mag as the floor
to the calculated errors, based on the observed dispersion in the transformation
between instrumental and standard magnitudes of bright stars (see below).
To derive the transformation of the instrumental magnitudes into the standard system
we used the following \citep{harris81}:

\begin{equation}
u'~=~u~-~k_u~x_u~+~ct_u~(u~-~g)~+~zp_u
\label{u_eq}
\end{equation}
\begin{equation}
g'~=~g~-~k_g~x_g~+~ct_g~(g~-~r)~+~zp_g
\label{g_eq}
\end{equation}
\begin{equation}
r'~=~r~-~k_r~x_r~+~ct_r~(r~-~i)~+~zp_r
\label{r_eq}
\end{equation}
\begin{equation}
i'~=~i~-~k_i~x_i~+~ct_i~(r~-~i)~+~zp_i
\label{i_eq}
\end{equation}
\begin{equation}
B~=~b~-~k_b~x_b~+~ct_b~(b~-~v)~+~zp_b
\label{B_eq}
\end{equation}
\begin{equation}
V~=~v~-~k_v~x_v~+~ct_v~(v~-~i)~+~zp_v.
\label{V_eq}
\end{equation}

\noindent In these equations $u'g'r'i'BV$ (left-hand side) are the published
magnitudes in the standard system \citep{landolt92,smith02}, $ugribv$ 
(right-hand side) correspond to the natural system magnitudes, $k_i$ is the
extinction coefficient, $x_i$ the effective airmass, $ct_i$ the color
term, and $zp_i$ the zero-point for filter $i$.

At the beginning of the first campaign, we observed a standards field hourly
($\sim$20 stars during the night), which allowed us to solve for all unknowns
($ct_i$, $zp_i$, $k_i$). As the campaign progressed the pressure
to observe SNe increased, leaving less time for measurements of standards.
We adopted the approach of measuring only 5--8 standard stars over a wide
range of airmass ($\sim$10 stars during the night). With this small
number of stars we fixed the color term to the average value, solving only 
for the extinction coefficient and zero-point.

\begin{figure}
\plotone{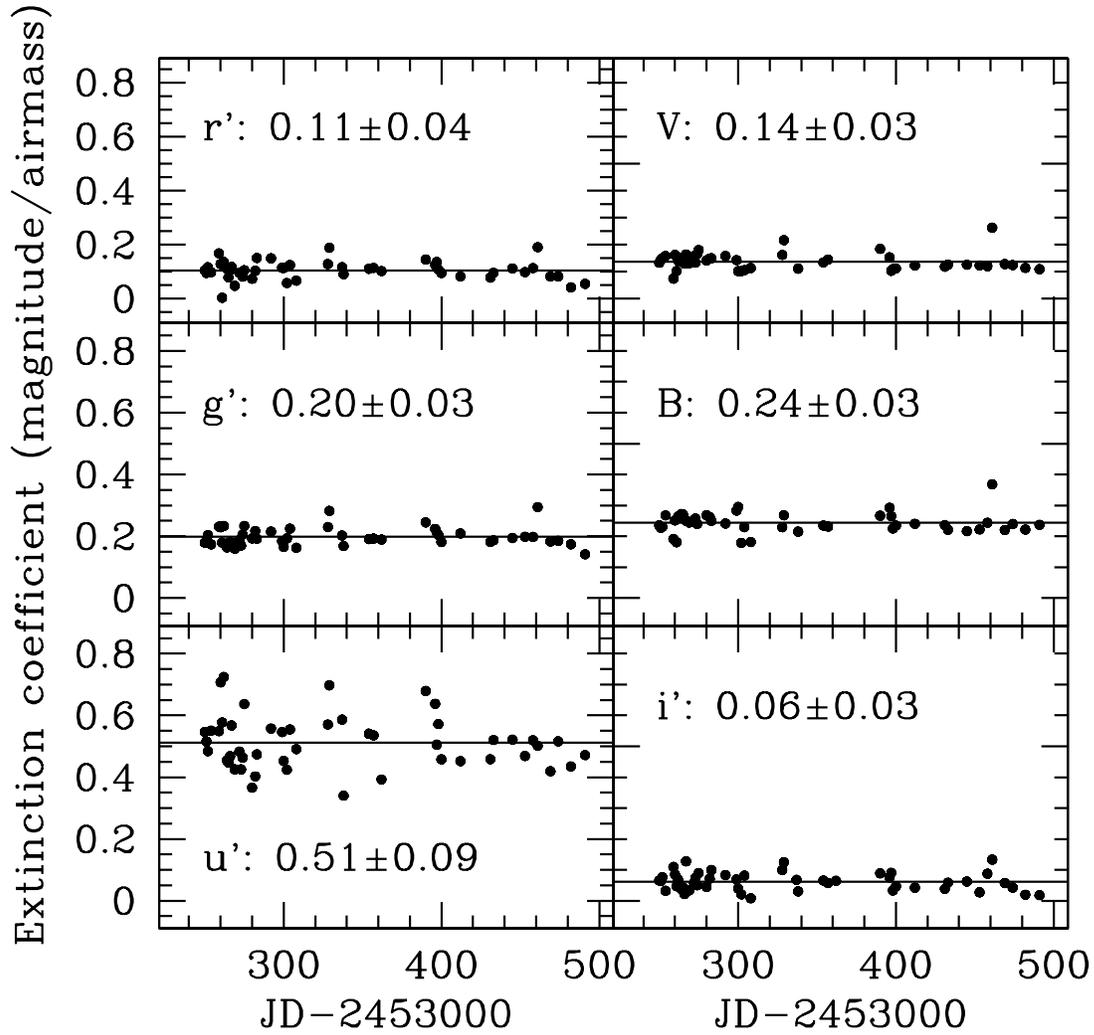}
\caption{Extinction coefficients for the $u'g'r'i'BV$ filters as a function of time. The average coefficient
is plotted with a horizontal line and its numerical value is indicated in each panel along with
the rms scatter.
\label{extinct_fig}}
\end{figure}

During the first observing campaign, we obtained weighted least-squares solutions
for 53 clear nights. Dispersions around best-fit transformations were
$\sim$0.02--0.04 mag in $u'$ and $\sim$0.01--0.02 mag in other filters.  
This clearly demonstrates the excellent quality of the LCO site for this 
photometric program.
Figure \ref{extinct_fig} shows the extinction coefficients as a function of time
for all filters. While the scatter was only $\sim$0.03 in $g'r'i'BV$, the
dispersion was three times greater in the $u'$ band which is the most sensitive
to extinction variations from night to night.

\begin{figure}
\plotone{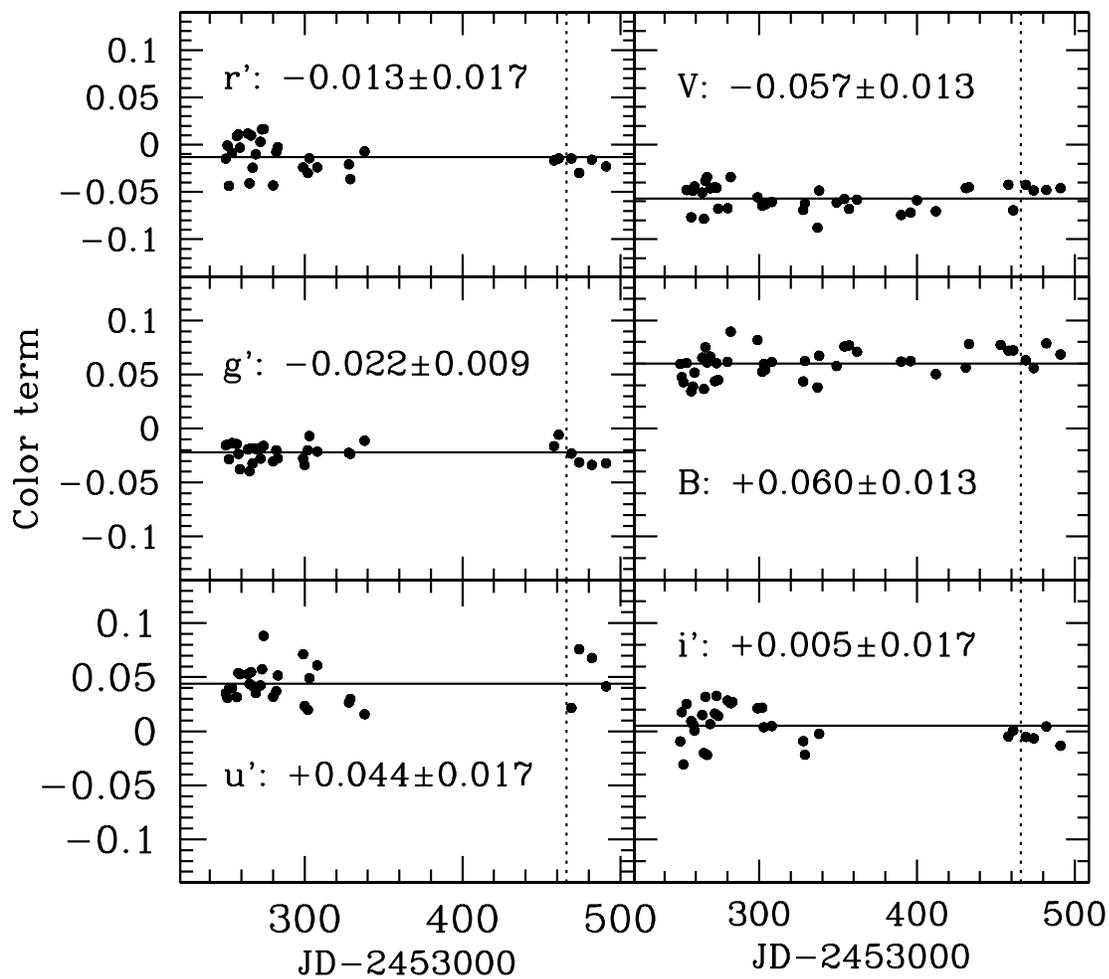}
\caption{Color terms for the $u'g'r'i'BV$ filters as a function of time. The average color term
is plotted with a horizontal line and its numerical value is indicated in each panel along with
the rms scatter. The vertical dotted line shows the time when the primary mirror of the 1~m telescope
was washed.
\label{pcoef_fig}}
\end{figure}

Color terms are presented in Figure \ref{pcoef_fig}. No obvious secular change
was seen over this 250-day span in any filter, even though the primary mirror
was washed on 2005 Apr. 06 UT. While the color terms were close to zero 
($\lesssim$0.02) for $g'r'i'$, the magnitudes of these terms in $u'BV$ were
greater ($\sim$0.04--0.06), indicating small, but non-negligible
differences between the instrumental and standard system.

To calibrate the local standards around the SN, we started by selecting
the 5--10 brightest stars in every SN frame. For each frame, we used 
these stars to compute aperture photometry and derive the magnitude correction
between a small aperture (generally $3''$) and the $7''$
aperture employed for the Johnson+SDSS standards.  The aperture
correction was typically $\sim$0.03--0.1 mag. We assumed there were no
field effects in the aperture corrections. Then we computed photometry
for 10--15 field stars through the small aperture and
corrected to the large aperture, bringing the natural system photometry to
the $7''$ radius aperture. While this correction added some complication
to the procedure, it improved the statistical accuracy of the fainter stars. 

Finally, we used equations \ref{u_eq}--\ref{V_eq} to derive magnitudes
in the standard system for the local standards. The uncertainty in the
resulting magnitudes was the Poisson error in the instrumental magnitudes
(assuming a minimum error of 0.015 mag). Here we neglected the uncertainty in the
aperture correction which was always less than 0.01 mag (otherwise we excluded such
measurements). From the multiple measurements obtained on different clear nights
we took weighted averages, yielding a photometric sequence of secondary standards
around each SN with uncertainties as small as 0.004 mag in the individual magnitudes. 

We then performed differential photometry of the SN relative to the local 
standards. The great advantage of this approach is that, since the SN is
observed simultaneously with the field stars, the magnitude differences
within a CCD frame are, to first order, immune to the passage of clouds.
To improve the instrumental precision, we performed point-spread-function (PSF) 
photometry. For this purpose, we employed all the stars of the photometric 
sequence to determine an average PSF for every CCD image, 
and we fitted the resulting PSF to 
the SN and the standards to a radius of $3''$. We converted the instrumental 
magnitudes to the standard system using equations \ref{u_eq}--\ref{V_eq},
assuming that the extinction effects ($k_i~x_i$) were simple additive 
constants which were absorbed, to first order, by the zero-point. As shown
in Figure \ref{pcoef_fig}, the color terms had no secular variations and
we adopted average color terms, solving only for the photometric zero-points.
The final uncertainty in the SN magnitude was the instrumental error in the PSF fit
(assuming a minimum of 0.015 mag, as explained above). We neglected errors due to 
the transformation to the standard system since the uncertainty
in the color term and in the zero-point are well below 0.015 mag.

The extraction of the SN magnitude from a CCD frame is generally uncertain
because the SN resides in a host galaxy with unknown features at
the SN position. Thus, the light curves computed so far should be considered
preliminary. As soon as we obtain galaxy templates with the 2.5~m telescope,
we will use image subtraction to remove the image of the galaxy and obtain
definitive light curves. This procedure will follow that described by 
\citet{hamuy94a}
which consists of (1) using the task ``geomap'' to determine the coordinate 
transformation 
between the two images  (assuming a two-dimensional linear polynomial) and 
using the task ``geotran'' to register the template to 
the SN+galaxy image; (2) using the task ``psfmatch'' to 
find the two-dimensional difference
kernel that, when convolved with the template, matches the PSF
of the SN+galaxy image; (3) using the task ``linmatch'' to match 
the flux scale of the template to
that of the SN+galaxy image; (4) subtracting the modified template from
the SN+galaxy image; and (5) extracting a small box around the SN from
the subtracted image and inserting it in the original SN+galaxy
image. 

The custom script package developed for the Swope telescope was used also
to process the WFCCD and LDSS-2 data through the flat-field division.
We have not attempted yet to compute SN magnitudes for these instruments
because the lack of standard observations has prevented us from
calculating color terms. This will be remedied during the second campaign.

\subsection{Infrared Imaging}
\label{nir_red}

We reduced both WIRC and PANIC data using software pipelines. These
pipelines are a combination of IRAF scripts following the steps of
(1) linearity correction, (2) dark combination and subtraction, (3)
flat-field combination and division, (4) bad-pixel mask production,
(5) sky image computation and subtraction, and (6) combination of
dithered frames into final stacked images. Both pipelines work in a
similar fashion, with slight differences due to differing observing
procedures.

In the first step, we applied a predetermined linearity correction law
for the HAWAII-1 detectors to every pixel value above
16,000 e$^-$: 

\begin{equation}
I_{corr} = c_0 + c_1~I + c_2~I^2,
\label{eq:IRlin}
\end{equation}

\noindent where $I$ is the observed number of ADU,
$I_{corr}$ is the corrected value, and the coefficients are set to 
$c_0 = 4.291443$E$+01$, $c_1 = 9.752524$E$-01$, and $c_2 = 1.962545$E$-06$. 
Next, we normalized the dark-subtracted dome flats by the median. We considered
pixel values outside the range 0.6--1.67 to be unrealistic, replacing them
by 1 in the normalized flat and marking them in the nightly bad-pixel mask.
If sky flats were taken (as was usually the case for PANIC), we median combined
them and normalized the result to an average value of 1.0, replacing pixel
values outside the range 0.2--5 by 1. We checked the results for possible 
residual stars. In the case of WIRC, we preferred sky flats (when available) 
to dome flats and we used them to check the illumination quality of the
latter, which was usually good to within 1\% for all filters. In the case
of PANIC, dome flats showed illumination gradients of about 5\% to 10\%
across the image. Therefore, when sky flats could not be obtained, we
used ones from a previous night.  We stored all combined darks, flats,
and bad-pixel masks of a night for future use or examination.

Sky frames to be subtracted from individual object frames were then
constructed. We combined off-target images to produce sky frames in
a two-step process. First, we computed a sky frame by directly
averaging all off-target images. We subtracted this first sky frame 
from the individual off-target images. We detected objects in
the subtracted frames and masked them in the original frames. We
recalculated  the sky frame, scaling the images to a common mode. We scaled
the sky to the mode of the individual (dithered) object frames and 
subtracted it from them. We computed also the mode of the object frame in
a two-step process to eliminate the contribution of sources in the
image. We obtained the final image by aligning and averaging the
individual object frames. In this process, we preserved all pixels 
except those marked in the bad-pixel mask.  Because the images may
have been taken in non-photometric conditions, we trimmed the data to
only the regions of complete overlap. We generally obtained two images
per SN and filter with WIRC (one from Detector \#2, and another from Detector
\#3), and one image with PANIC.

Once the pipelines had produced stacked science images, we proceeded with
the photometric measurements. Every time standard stars were observed with WIRC,
we measured instrumental magnitudes through a standard aperture of $5''$ in
radius \citep{persson98}, with a sky annulus at $5''$ to $7''$ from the
star. The magnitudes carried associated statistical uncertainties based
on a Poisson model of the noise. As with the optical photometry, these
estimates turned out to be unrealistically small when the objects were
bright. Based on the dispersion found in the photometric
solutions (see below), we estimate the minimum error in a 
single measurement to be 0.02 mag. This was our adopted minimum
uncertainty in the instrumental magnitudes.

We used the instrumental magnitudes of standard stars to solve for the
photometric transformation of the night, defined by the following
equations:

\begin{equation}
Y = y - k_y~x_y + zp_y,
\label{Y_eq}
\end{equation}
\begin{equation}
J = j - k_j~x_j + zp_j,
\label{J_eq}
\end{equation}
\begin{equation}
H = h - k_h~x_h + zp_h,
\label{H_eq}
\end{equation}
\begin{equation}
K_s = k - k_k~x_k + zp_k.
\label{K_eq}
\end{equation}

\noindent Here, $YJHK_s$ are the magnitudes in the standard system
(the $JHK_s$ magnitudes published by \cite{persson98} and the $Y$ magnitudes
given in appendix \ref{yband_app}), $yjhk$ are the corresponding instrumental magnitudes, 
$k_i$ is the extinction coefficient, $x_i$ the effective airmass, and $zp_i$ the
zero-point for filter $i$. We fixed the extinction coefficients for $JHK_s$
to the canonical values given by \citet{persson98}: $k_j=0.1$, $k_h=0.05$,
and $k_k=0.08$. In the case of $Y$, we used the same value as for $J$
($k_y=0.1$). We employed a 
weighted least-squares method to find the zero-points $zp_i$. 
Typical dispersions around the fits were $\sim$0.02--0.03 mag for all filters.
On six nights, we observed 4--5 standards spread over airmass 1--2
in order to also fit the extinction coefficients. We found no
significant deviations from the nominal values for $JHK_s$. In the $Y$ band
we found an average of $k_y=0.06\pm0.01$, in agreement with the value of 
$k_y=0.047$ given by \citet{hillenbrand02}.

Note that we assumed zero color terms since the instrument detector and
filters were essentially the same as in \citet{persson98}. 
We will verify and monitor this assumption during the course of future campaigns.

With the photometric solution for the night, we proceeded to calibrate the local
standard stars in each SN field observed with Detector \#2 of WIRC. We started
by measuring magnitudes of bright, isolated stars through apertures of
$1''$ to $5''$ in every stacked science frame and deriving the magnitude
correction between a $2''$ and a $5''$ radius aperture. Typical aperture
corrections ranged between 0.02 and 0.1 mag.  We kept this
correction below 0.1 mag by increasing the size of the small aperture,
if necessary.  Since the uncertainty in the aperture correction was
generally $\sim$0.02 mag, the increase in the statistical
uncertainty on the $2''$ measured magnitude was marginal and
certainly lower than that introduced by increasing the aperture to $5''$. 
We measured magnitudes for a number of comparison stars (between 3 and 15)
in each SN field through the small aperture and corrected them to the
$5''$ aperture. We transformed these instrumental magnitudes to
the standard system using equations (\ref{Y_eq}) to (\ref{K_eq}). 
The weighted averaged of the resulting magnitudes for all comparison stars 
obtained on different photometric nights was calculated and checked 
for consistency.  We then computed SN magnitudes relative to comparison 
stars in a manner identical to that used for the optical data.

\subsection{Spectroscopy}

 The first step in the spectroscopic reductions was to combine the
bias and flat-field images, and then process the science frames
through overscan subtraction, trimming, bias correction, and
flat-fielding.  Next, a general wavelength calibration was derived for
the night, which we used later as a starting point to obtain specific
calibrations from the comparison lamp images obtained at the position
of each SN. Our wavelength calibrations were typically characterized
by a root-mean-square (rms) scatter of 0.1 pixel.

The sensitivity function for the night was then calculated. This
procedure consisted of extracting 1-D spectra of the flux standards,
applying the wavelength calibration, dividing them by the 1-D telluric
standard spectrum (after eliminating by hand the weak intrinsic
spectral features of such stars), and calculating a response curve
based on the flux values measured for the standard stars.  As pointed
out by \cite{bessell90} (see also Matheson et al. 2000), 
the division by the telluric spectrum has the
following advantages: (1) it effectively removes the high-frequency
(10--250 pixel wavelength) wiggles in the spectrum that are introduced
by typical flat-field continuum sources and which can have a higher
frequency than the flux point spacing, especially in the red, (2) the
response curve can be fit accurately with a low-order function, and (3)
most of the telluric features disappear, except those which are
strongly saturated (like the oxygen band near 7,600 \AA). The latter
effect is very important in revealing the calcium, oxygen, and carbon
features in the red.

Extraction of the SN spectra was accomplished using a window
whose width was chosen
depending on the seeing and the
uniformity of the underlying galaxy distribution. Typically the window included 
$\sim$90\% of the SN flux and excluded the wings of the stellar profile where
the contribution of the galaxy introduced unwanted noise. We used two adjacent
windows along the spatial direction to estimate the sky level and the contribution of the
galaxy in the SN aperture. In general we tried to select these background windows
as close as possible to the SN aperture, but the criteria for their selection
varied from case to case depending on the nature of the adjacent
background. A cubic spline was used
to interpolate the background at the SN position.
In most cases this interpolation provided an accurate estimate of the actual 
background, but occasionally the background was so non-uniform that some 
contamination of the host galaxy in the 1-D SN spectrum was unavoidable. 

The extracted 1-D SN spectrum was then wavelength calibrated using the 
comparison lamp exposure taken at the SN position.  Finally, 
the wavelength-calibrated spectra were divided by the 1-D telluric spectrum, 
and the sensitivity function was applied.  Multiple observations of the same
SN were combined into a final spectrum using a median-filter algorithm to 
remove deviant pixels caused by cosmic rays. 

\section{FIRST RESULTS}
\label{res}

During the first low-$z$ CSP campaign, we observed a total of 72 SNe. However,
only 38 of these (17 SNe~Ia, 12 SNe~II, and 9 SNe~Ibc) eventually qualified 
for inclusion in our follow-up program. This number represents 76\% of our
nominal expectation of $\sim$50 SNe, with the difference being ascribed to
natural variations in the year-to-year SN discovery rates.

\subsection{Optical Imaging}

During the first campaign, we obtained 7852 science optical images,
and established photometric sequences around all of the 38 SNe
included in our follow-up observations. Final light curves must await
the eventual subtraction of template images, but we present here
preliminary light curves obtained with the Swope telescope of two SNe
that did not suffer significant contamination from their hosts. While
we chose these SNe for minimal contamination, the effect is bound to
appear at some point when the SNe became very faint.  Nevertheless,
these examples illustrate the excellent photometric quality of the CSP
light curves.

\begin{figure}
\plotone{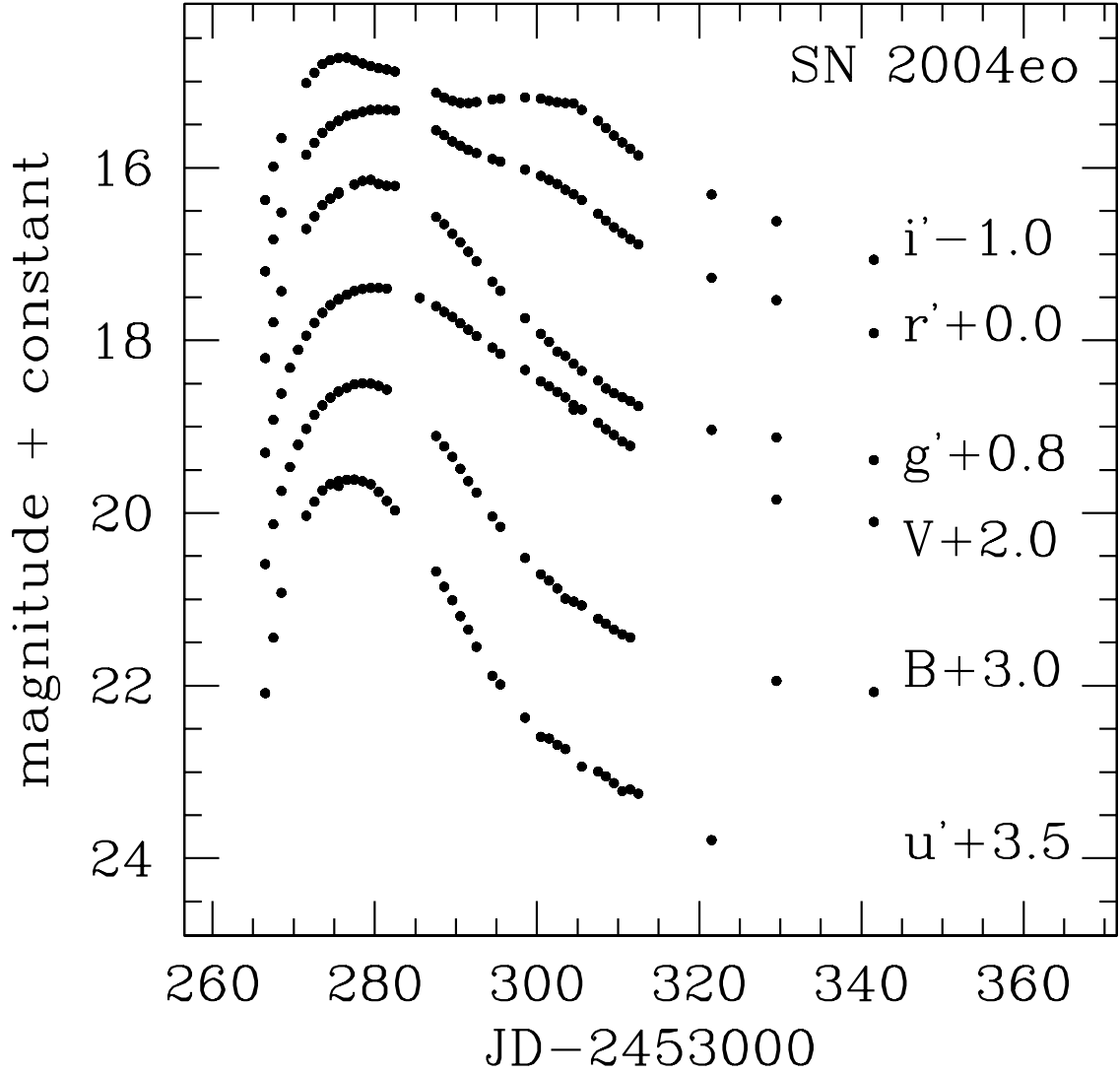}
\caption{$u'g'r'i'BV$ light curves of the Type Ia SN~2004eo.
\label{sn04eo_opt_fig}}
\end{figure}

Figure \ref{sn04eo_opt_fig} shows the $u'g'r'i'BV$ light curves of the
Type~\,Ia\,~SN\,~2004eo. We observed this object for $\sim$80 days
starting $\sim$10 days before maximum light with a cadence rarely
achieved in previous studies. To estimate the precision of our
photometry we fit a L\'egendre polynomial to the light curves with the
lowest possible order but making sure to eliminate systematic
residuals. In this case, the scatter around the fits amounts to 0.028
(order 8), 0.012 (order 9), 0.006 (order 12), 0.012 (order 15), 0.010
(order 9), and 0.008 (order 10) mag in $u', g', r', i', B$, and $V$,
respectively.  These dispersions can be taken as an empirical and
realistic estimate of the random error in a single observation.

\begin{figure}
\plotone{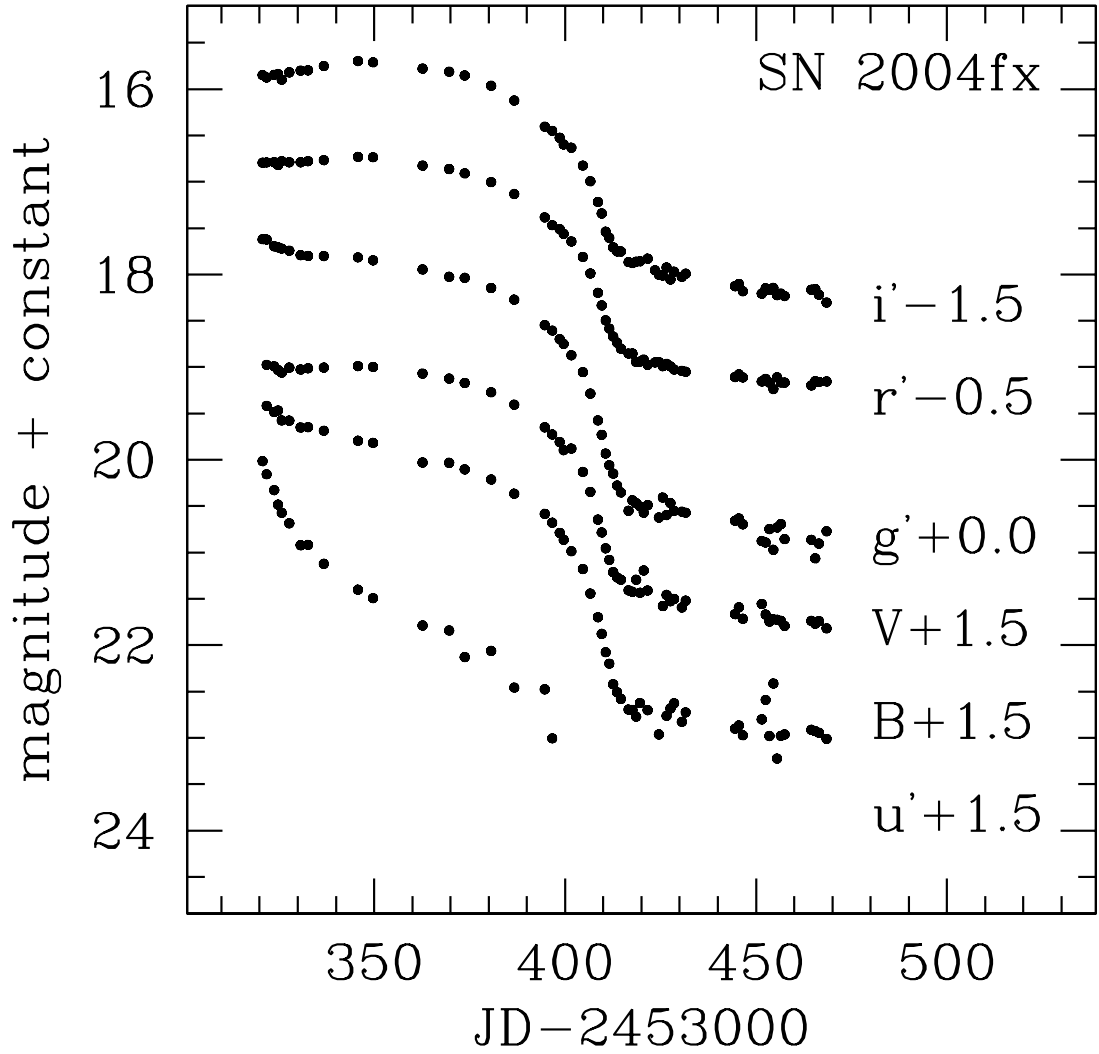}
\caption{$u'g'r'i'BV$ light curves of the Type II-P SN~2004fx.
\label{sn04fx_opt_fig}}
\end{figure}

In Figure \ref{sn04fx_opt_fig} we present optical light curves for the
Type~II plateau SN\,~2004fx.  A pre-discovery image on 2004 Oct. 22,
where the SN was not visible, indicates that the SN was caught no
later than 20 days after explosion. Our first observations confirm
that the SN was still quite blue at that point and that it grew
progressively redder. We observed the SN on a weekly basis through the
plateau phase for $\sim$80 days, every other night for the next 20
days as it evolved faster, and less frequently again during the linear
phase. The scatter in $u'$ about a fourth-order L\'egendre polynomial
is 0.10 mag, but this is dominated by the last few points where the
statistical uncertainties are much greater. When the four points with
the greatest residuals are removed, the scatter drops to 0.044 mag. In
the other bands, the scatter during the plateau phase around an order
10 L\'egendre polynomial varies between 0.02 and 0.03 mag. During the
late-time linear phase the dispersion around a linear fit varies
between 0.03 and 0.10 mag in $g', r', i', B, V$.  Part of this scatter
is caused by variable amounts of galaxy contamination as the seeing
varied during the observations. We expect the final light curves to be
much more uniform at these late epochs.

\subsection{Infrared Imaging}

During the first campaign, we obtained thousands of individual NIR
images. The reduction process produced 1449 calibrated mosaics.  We
established photometric sequences around all of the 24 SNe included in
our follow-up observations. Our NIR light curves are preliminary,
awaiting the final template images and subtraction.

\begin{figure}
\plotone{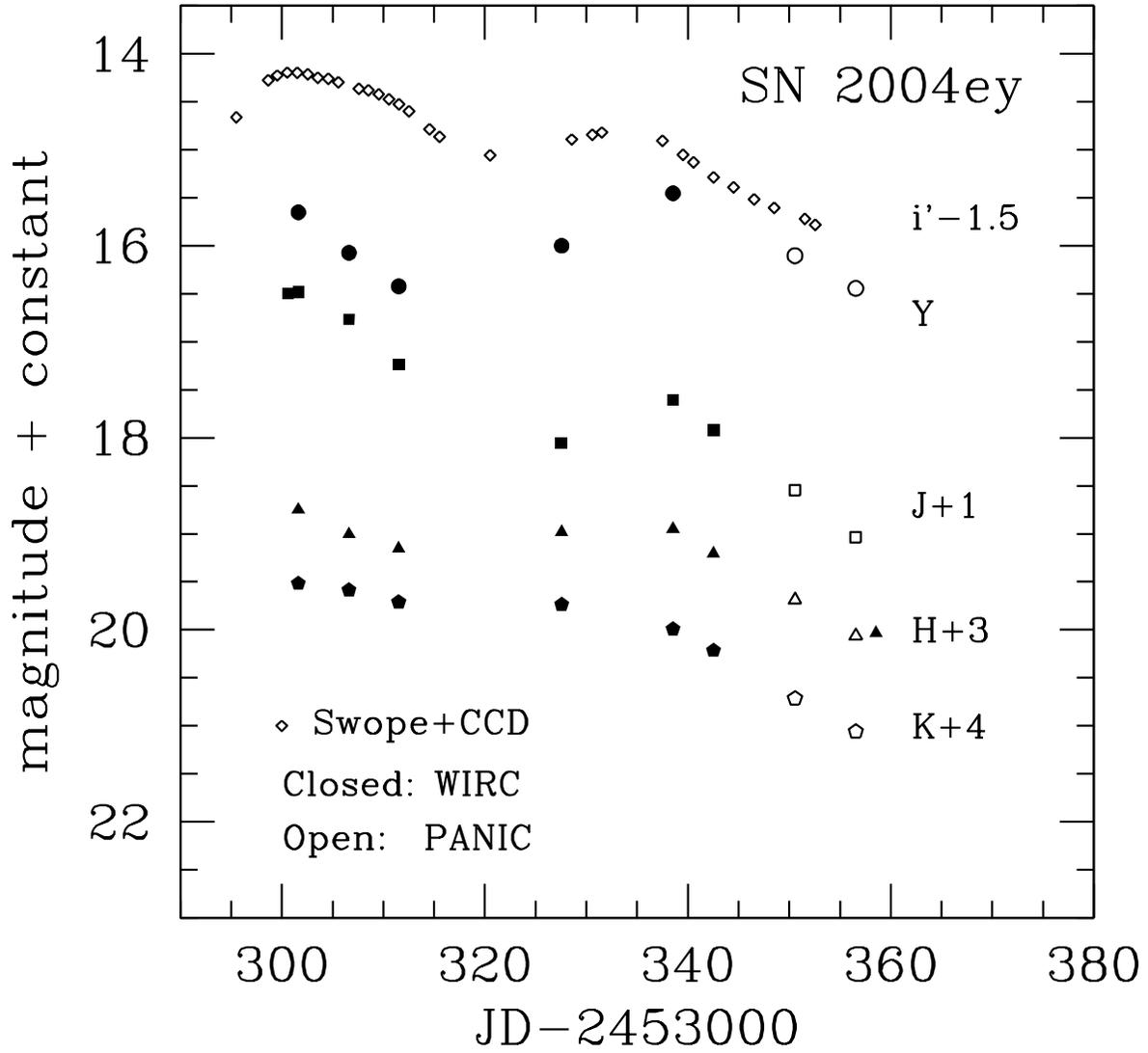}
\caption{$YJHK_s$ light curves of the Type Ia SN~2004ey. The $i'$-band light curve is also
shown to guide the eye.
\label{sn04ey_IR_fig}}
\end{figure}

\begin{figure}
\plotone{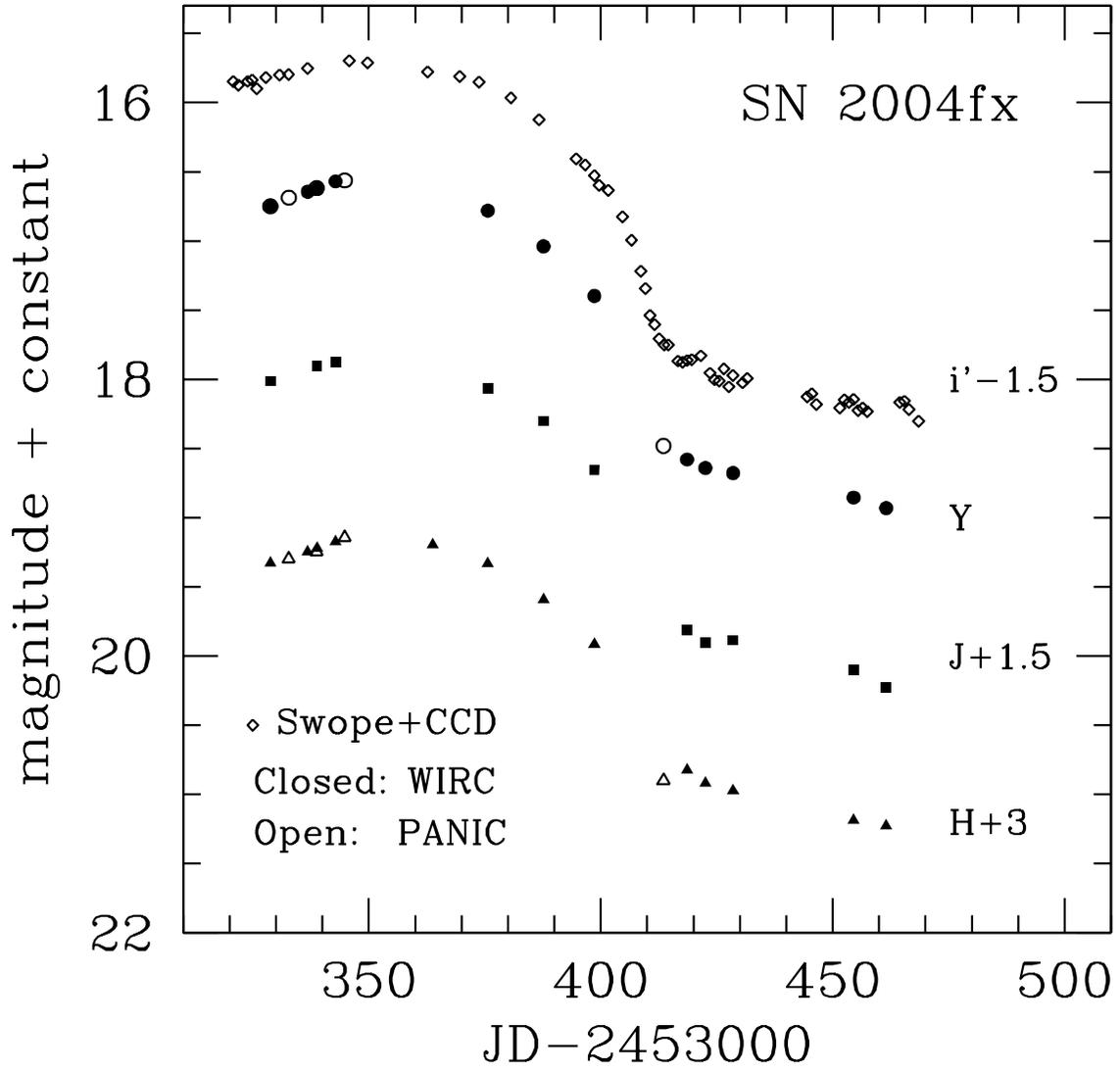}
\caption{$YJH$ light curves of the Type II-P SN~2004fx. The $i'$-band light curve is also
shown to guide the eye.
\label{sn04fx_IR_fig}}
\end{figure}

The NIR photometry is presented in the $YJHK_s$ light curves
for two representative cases. Figure \ref{sn04ey_IR_fig} presents the
Type~Ia SN~2004ey, while Figure \ref{sn04fx_IR_fig} shows the
Type~II SN~2004fx. In both cases, 
we show for comparison the $i'$-band light curves
from our own optical follow-up program.

The light curves show the typical sampling achieved of one point every
5--7 days, with gaps of up to 15 days during lunar ``dark runs'' when WIRC was
not available at the 2.5~m duPont telescope. By exchanging time with
other programs, we were able to improve the sampling of some
light curves, especially in the cases of SNe~Ia around maximum
light, to one point every 1--2 days. For SNe~II, we relaxed the
frequency of observations to one point every $\sim$10 days,
enough to follow the slow plateau evolution.  From the observations reduced
independently for Detectors \#2 and \#3 of WIRC, we are able to estimate
the precision of the measurements.  In cases of high
S/N, where photon uncertainties can be neglected,
the deviation between the two points is $\sim$0.02--0.03 mag in the
$YJH$ bands, consistent with expectations.

The distinctive secondary maximum of SNe~Ia is clearly seen in
Figure \ref{sn04ey_IR_fig}, occurring about 20 days after first
maximum. This secondary maximum is remarkably prominent in $Y$, even
surpassing the brightness of the first maximum.  The NIR behavior of
the Type~\,II\,~SN is characterized by a slow yet steady luminosity
increase during the plateau phase, a feature previously seen in SN\,~1999em \citep{hamuy01}.

\subsection{Spectroscopy}

During the first CSP campaign, we obtained a total of 213 optical
spectra which are fully reduced. During this period we provided
spectral classification in the IAU Circulars for 27 SNe.

\begin{figure}
\includegraphics[angle=-90,width=18cm]{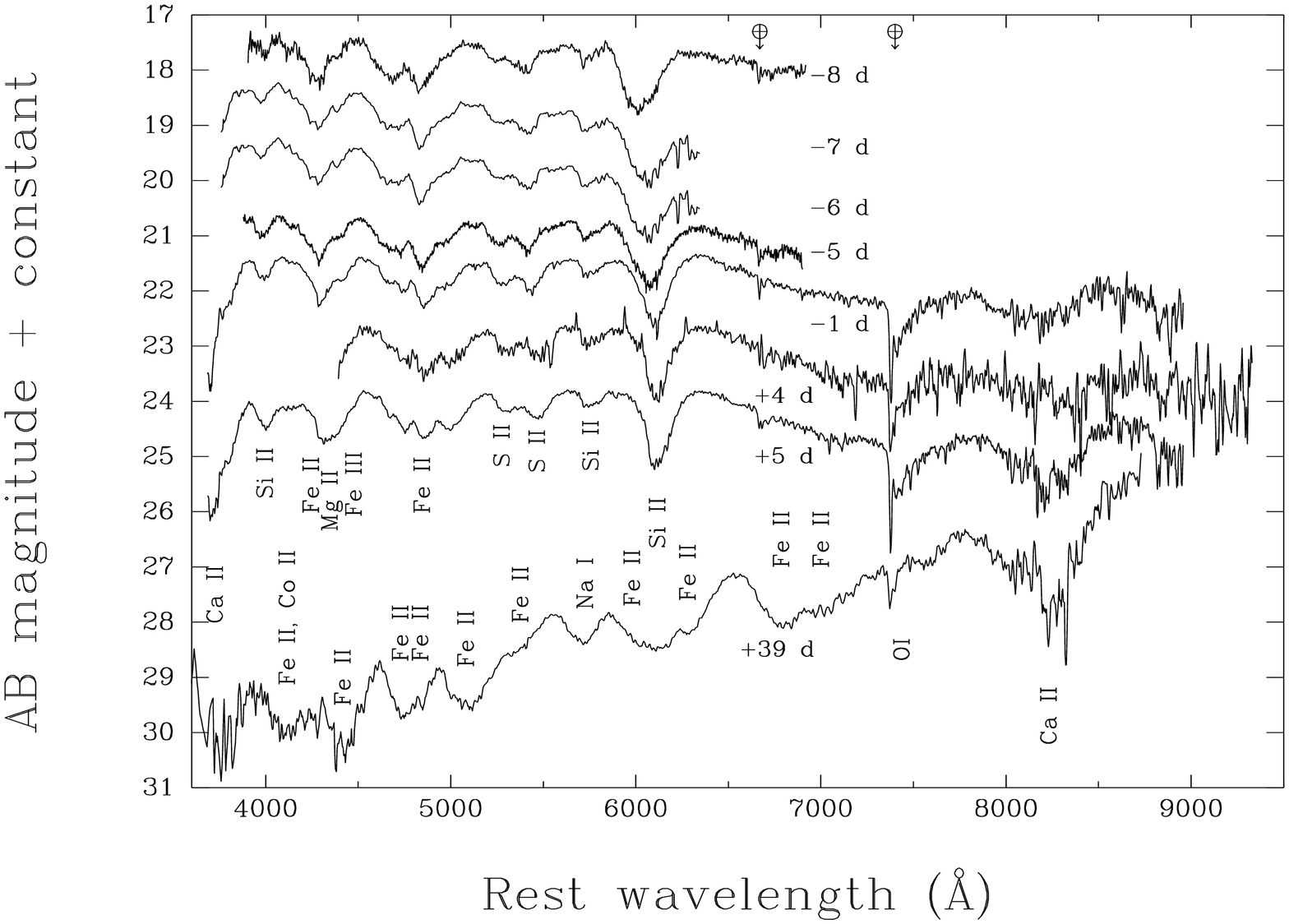}
\caption{Spectroscopic evolution of the Type~Ia SN~2004ef. The telluric features
are indicated with the $\earth$ symbol. AB mag = $-2.5\, {\rm log}\, f_\nu - 48.6$ \citep{oke83}.}
\label{sn04ef_z}
\end{figure}

Figure \ref{sn04ef_z} displays the spectroscopic evolution of the 
Type~\,Ia\,~SN\,~2004ef starting 8 days before maximum light for a period of 47
days. The telluric features are
quite evident in these spectra because we did not obtain telluric standard
observations for this SN.

\begin{figure}
\includegraphics[angle=-90,width=18cm]{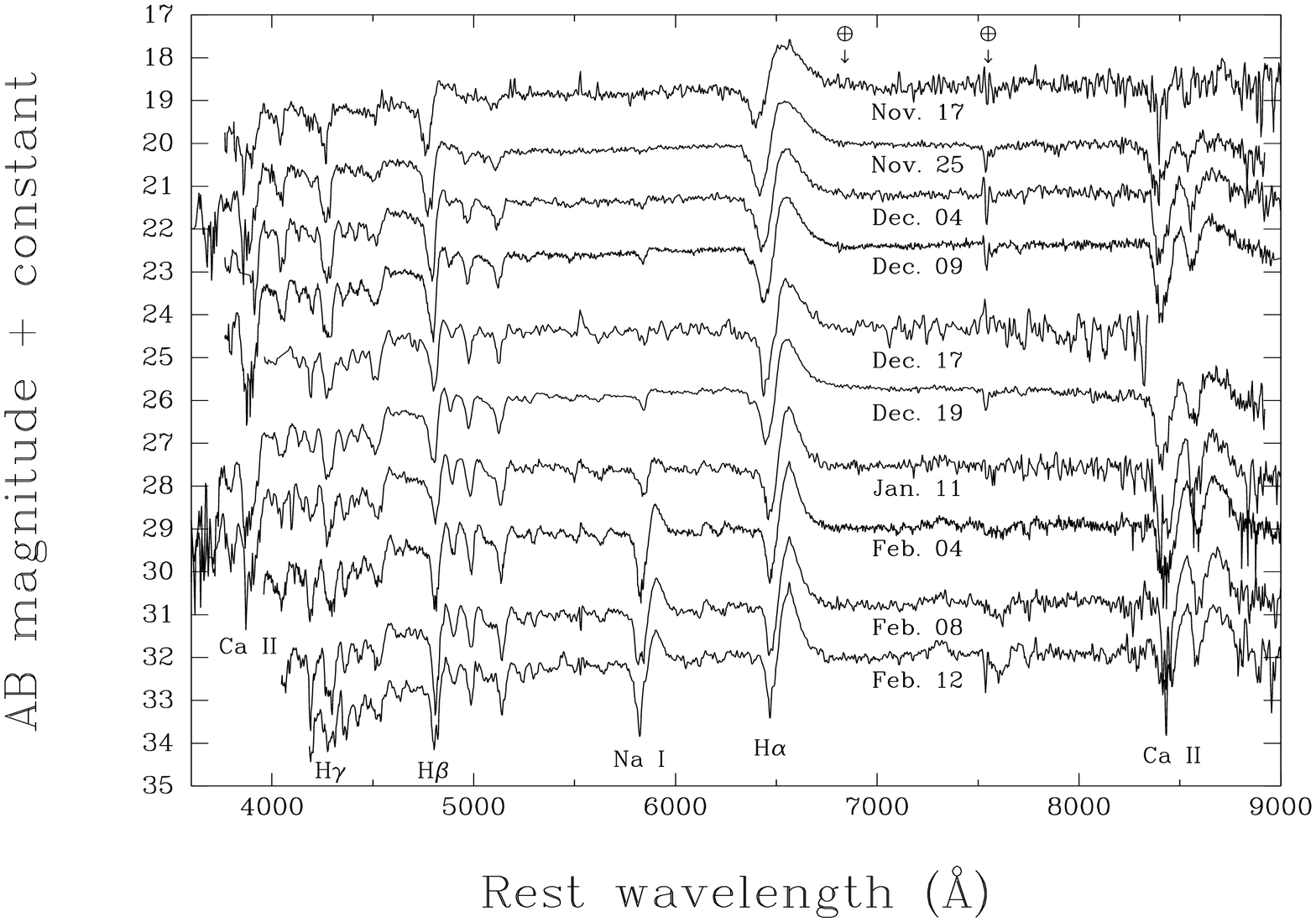}
\caption{Spectroscopic evolution of the Type~II-P SN~2004fx. The telluric features are indicated with the $\earth$ symbol. AB mag = $-2.5\, {\rm log}\, f_\nu - 48.6$ \citep{oke83}.}
\label{sn04fx_z}
\end{figure}

Figure \ref{sn04fx_z} shows the temporal evolution of the spectrum of the
Type~II SN~2004fx. Unlike the case for SN~2004ef, 
these spectra were divided by a telluric
standard. Most telluric features disappear and only small residuals can be 
seen at the strongest absorptions (like the ``A band'' near 7,600~\AA).


\section{FUTURE IMPROVEMENTS TO THE CSP} 
\label{future_sec}

In this paper, we have outlined the basic strategy of the CSP. In the
second campaign, we will be modifying small aspects to improve the data 
quality still further.

The most important addition to the CSP will be RetroCam, 
a simple, high-throughput, NIR
imager that will be mounted continuously on the Swope
telescope during each SN campaign. It has a 
HAWAII-1  $1024 \times 1024$ pixel HgCdTe detector, and a single filter wheel
containing $Y$, $J$, and $H$ filters. Because it has no
reimaging optics it does not operate in the $K_s$ band.
The scale is $0.54''$ pixel$^{-1}$, which is
adequate for the image quality
delivered by the telescope. 
RetroCam and the CCD camera (used for all the
Swope measurements made so far) will both be
mounted on a mechanical ``swivel,'' such that they can
be exchanged with each other in a matter of minutes.
Typically this will be done during the afternoon,
depending on SN scheduling requirements, weather,
and so forth. Thus we expect to be able to alternate
between $u'g'r'i'BV$ and $YJH$ photometry 
every two or three nights or, in exceptional
cases, during the same night. We expect that the final
result will be light curves with comparable density
of coverage from $u'$ through $H$, and with sparser $K_s$
coverage from WIRC and PANIC.

During the first campaign, we did not have enough engineering or cloudy
time on our WIRC or PANIC nights to allow us to perform linearity tests.
Hence, we have relied on the corrections that are given in the manuals of
these facility instruments.
We will improve this situation
by carrying out our own tests during the second campaign.

During the first campaign, it was not always possible to achieve 
an optimal cadence of one spectroscopic observation every
5--7 nights per SN.  
We expect to improve this situation for future CSP campaigns by obtaining
supplementary spectroscopic time on the
CTIO 4~m and the European Southern Observatory NTT telescopes.
Additional spectra will also be obtained at Lick Observatory.

\section{SUMMARY AND CONCLUSIONS}

The low-$z$ program of the CSP is underway. During the first 9-month
campaign, we were assigned 190 nights with the LCO 1~m telescope, 57
nights with the 2.5~m telescope, and a smaller number of nights with
the two Magellan 6.5~m telescopes and the CTIO 1.5~m telescope.  With this
allocation, we were able to obtain follow-up photometry and
spectroscopy for 38 SNe (17 SNe~Ia, 12 SNe~II, and 9 SNe~Ibc).

In the first year of operation, we developed reduction pipelines
and software which allowed us to produce optical+NIR light curves in
real time and post them on our public web site. Thanks to the nearly
uninterrupted access to the Swope 1~m telescope, our optical
light curves have an unprecedented gap-free temporal coverage.  This
data set constitutes the first ever for SNe in the SDSS filters. Given
that we used only one instrument and filter set, the light curves are
very homogeneous in comparison to data sets in the literature such as the
``gold sample'' of \cite{riess04}. Through careful attention to details,
precisions of 0.03 mag in
$u'$ and 0.01 mag in $g'r'i'BV$ in single measurements have been
achieved.
We were able to sample the NIR light curves of SNe~Ia every
5--7 days with longer gaps of 15 days during dark time, and those of 
SNe~II every 10
nights. The precision of single measurements in the $YJH$ bands was typically
0.02--0.03 mag, consistent with expectations.

Our data processing procedures allowed us to flux and wavelength calibrate
hundreds of spectra in a timely manner. During the first campaign, the CSP 
provided
spectroscopic types for 27 SNe.
These spectra will serve as a valuable resource for improving
K-corrections for SNe~Ia and SNe~II, as well as for
measuring expansion velocities and line strengths in order to explore
correlations with SN luminosities and thus refine the methods for
distance determination.

\acknowledgments

\noindent
We thank Janusz Kaluzny for providing linearity calibrations for the
Swope CCD.
This material is based upon work supported by the National Science
Foundation (NSF) under grant AST--0306969. 
We also acknowledge support from {\it Hubble Space Telescope}
grant GO-09860.07-A from
the Space Telescope Science Institute, which is operated by
the Association of Universities for Research in Astronomy, Inc., under 
NASA contract NAS 5-26555. MH acknowledges support 
provided by NASA through Hubble Fellowship grant HST-HF-01139.01-A,
and support from the Centro de Astrof\'\i sica FONDAP 15010003.
A.V.F.'s group at U.C. Berkeley is supported by NSF grant AST-0307894; he is
also grateful for a Miller Research Professorship, during which part of this
work was completed.  KAIT was made possible by generous donations from Sun
Microsystems, Inc., the Hewlett-Packard Company, AutoScope Corporation, Lick
Observatory, the National Science Foundation, the University of California, and
the Sylvia \& Jim Katzman Foundation.

\appendix

APPENDIXES

\section {SHUTTER CORRECTIONS}
\label{shutter_app}

For the curtain shutter design of the Swope 1~m telescope
CCD camera, we expect a single
constant error across the field. We attempted to measure this effect
during several cloudy nights. With the dome closed, we adjusted the flat
field lights until the illumination yielded 25,000 e$^-$~pixel$^{-1}$
in 30 s. The measurements consisted in taking one 30 s exposure, ten 3 s
exposures without reading the CCD in between the individual exposures,
and another single 30 s exposure. In the case of an additive shutter
error, $t_s$, the total effective exposure time for the two 30 s images
was 30+$t_s~$ s, while for the middle image the total effective exposure
time was 10$\times$(3+$t_s$) s. The illumination ratio between the
10$\times$3 s image and the average of the two 1$\times$30 s images, $r$,
was very close to one and we used it to solve for $t_s$ from

\begin{equation}
t_s~=~\frac {30~(r~-~1)} {10~-~r} .
\end{equation}

In October 2004, we determined $t_s$ with the telescope pointed to four 
different positions without detecting significant differences.
From 14 measurements we measured $t_s$=0.079$\pm$0.009 (rms). We repeated
this test four times during the campaign (with one telescope pointing), 
which yielded $t_s$=0.080$\pm$0.007, 0.075$\pm$0.009, 0.077$\pm$0.004, 
in agreement with the first test. On one night, we confirmed that $t_s$
was an additive constant, independent of the exposure time. For
this purpose, we took 10$\times$4 s exposures bracketed by two 1$\times$40 s
images which yielded $t_s$=0.083$\pm$0.003, and 10$\times$5 s exposures bracketed by
two 1$\times$50 s images which yielded $t_s$=0.084$\pm$0.003. These tests demonstrated
that the shutter error was important, especially for short exposure times.
For example, neglect of the correction would have introduced a 0.08 mag error in a 1 s
exposure relative to a long one. Note that the uncertainty in this correction
is only 0.002 mag in a 1 s exposure and proportionally 
less for longer exposures.

\section {LINEARITY CORRECTIONS}
\label{linearity_app}

When the sky was overcast, we used the Swope 1~m telescope to measure the
linearity of the SITe CCD. With the dome closed during the night, we pointed
the telescope to a white flat-field screen, 
adjusting the intensity of the quartz lamps to give 100 ADU~pixel$^{-1}$ above the bias in 8 s.
Then we took exposures of 4, 8, 16, 24, ... 1800 s, which allowed us
to measure the CCD response between 50 and 23,000 ADU~pixel$^{-1}$ (57,500 e$^-$~pixel$^{-1}$),
above which we noticed that the CCD became quite nonlinear. We bracketed each of
these images with 8 s images in order to monitor the illumination
drift of the quartz lamp. We found that the illumination could vary
by up to $\sim$2\% during the 2-hour time span of the measurement
sequence.  Once we corrected the fluxes for these illumination variations, we
fitted a model of the form

\begin{equation}
t_i+t_s = a_1~I_i~\left(1 + a_2 \frac{I_i}{32767} + a_3 \frac{I_i^2}{32767^2} \right),
\label{nl_eq}
\end{equation}

\noindent where $I_i$ was the flux detected in $t_i$ s. 

\begin{figure}
\plotone{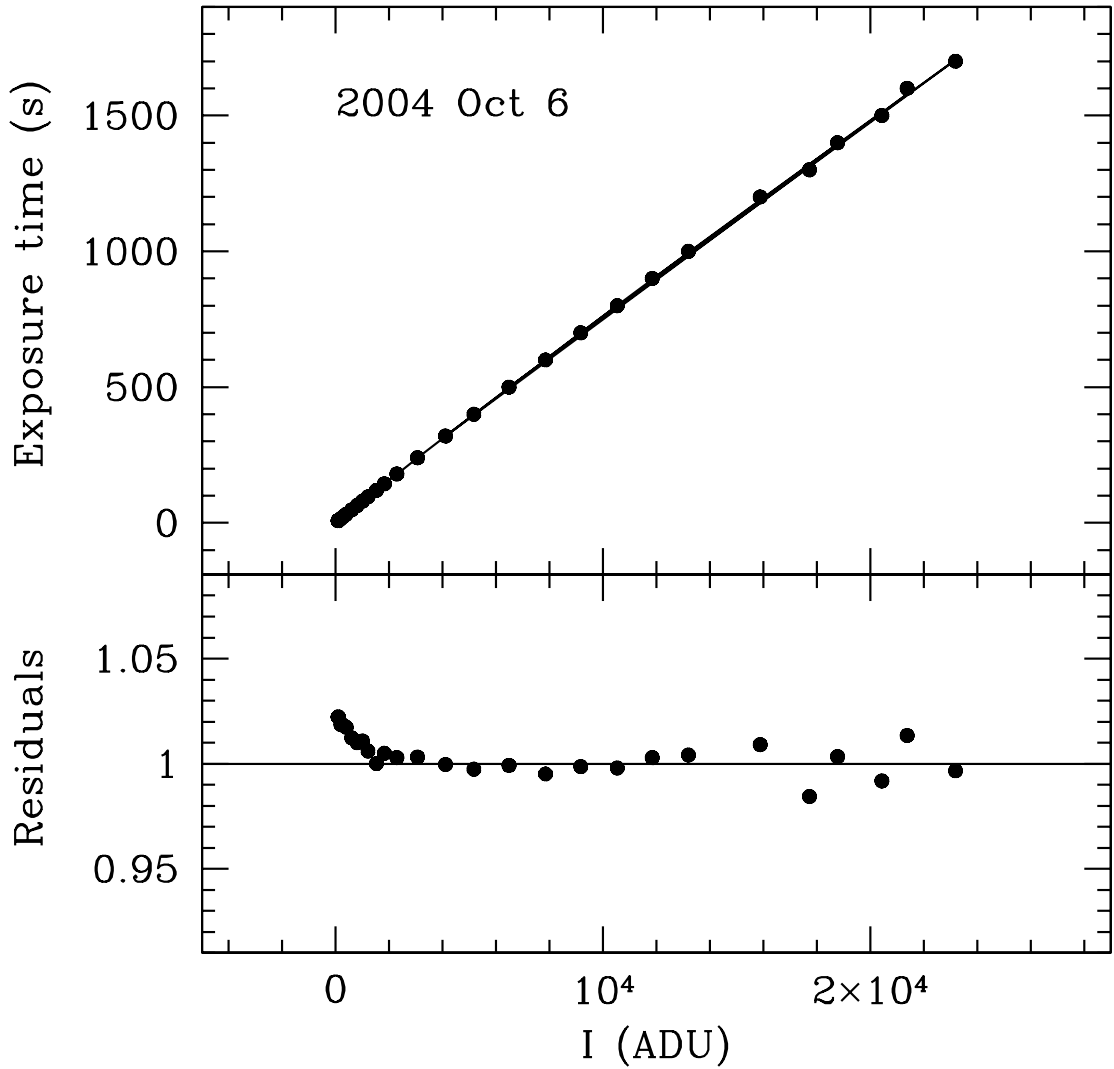}
\caption{(top) Illumination per pixel versus exposure time (corrected for shutter error) for the
Swope SITe CCD (filled circles), measured on 2004 Oct. 6. The solid line 
is a cubic polynomial model
(cf. equation \ref{nl_eq}). (bottom) Fractional residuals from the best fit.
\label{nl_fig}}
\end{figure}

Figure \ref{nl_fig} (top) shows measurements obtained on 2004 Oct. 6
(filled circles) and the best-fit model.  The bottom panel shows the
fractional residuals from the best fit. Although at low-illumination
levels the residuals amount to $\sim$2\%, the cubic polynomial
provides a reasonable fit to the CCD response over the whole dynamic
range. A better fit would have been obtained with a higher-order
polynomial, but we preferred to avoid an excessive number of free
parameters in the model\footnote{We have empirically measured that the
effect of adding a fourth-order term to the polynomial fit affects the
SN magnitudes by $\lesssim$ 0.002 mag, thus demonstrating that this
correction is unnecessary.}.  We then proceeded to linearize the
observed fluxes on a pixel-by-pixel basis using the formula

\begin{equation}
I_{corr} = I~\left(1 + a_2 \frac{I}{32767} + a_3 \frac{I^2}{32767^2} \right). 
\label{lc_eq}
\end{equation}

\begin{figure}
\plotone{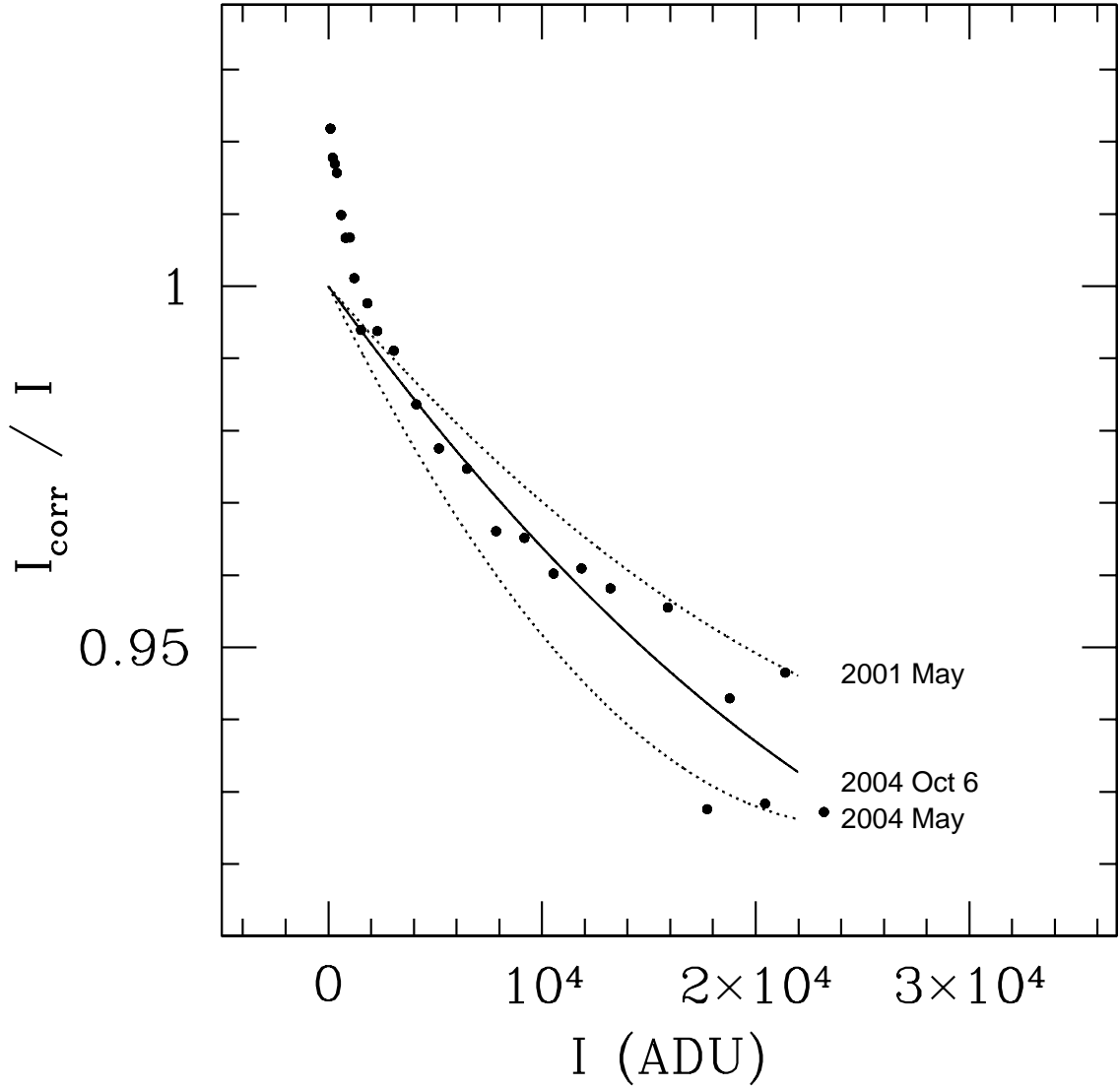}
\caption{Non-linear correction for the Swope SITe CCD as a function of the number of
detected ADU. The points show the data we obtained on 2004 Oct. 6
and the solid line shows the best-fit model. The two fits shown with dotted lines
were found by Janusz Kaluzny prior to the CSP (May 2001 and May 2004). These fits were
chosen to represent the full range of corrections that have been measured
for the SITe detector.
\label{lcoeff_fig}}
\end{figure}

 The last four rows of Table \ref{tab_coef} list the resulting coefficients from 
four different nights when we carried out these linearity measurements. 
Figure \ref{lcoeff_fig} shows the data obtained on 2004 Oct. 6 along with
the best-fit model (solid line).
Note that the departures from linearity reached almost 6\% at 23,000 ADU,
so it was absolutely necessary to correct the data for this effect to achieve
high-precision photometry. Above this value the curve shown in Figure \ref{lcoeff_fig} 
has a steep upturn, so we find it safer to avoid illuminations above 23,000 ADU
and define our saturation limit at this value.

   Included in Table \ref{tab_coef} are previous measurements
of the linearity of the SITe CCD carried out between May~2001 and May~2004
by Janusz Kaluzny.  Two of these measurements are plotted in  
Figure \ref{lcoeff_fig}.  Note that the fits shown in Figure \ref{lcoeff_fig} 
represent the full range of the linearity corrections that have been 
measured for the SITe CCD.  
These differences are almost certainly due to errors
in the measurements themselves rather than changes in the linearity of the
detector.  Hence, for the entire first campaign of the CSP, we adopted 
the correction measured on 2004 Oct. 6, which lies in the middle of the 
range of fits.  The uncertainty in this linearity correction 
translates to a maximum error in the photometry which is less than 2\%.

\section{THE Y-BAND PHOTOMETRIC SYSTEM}
\label{yband_app}

\begin{figure}
\plotone{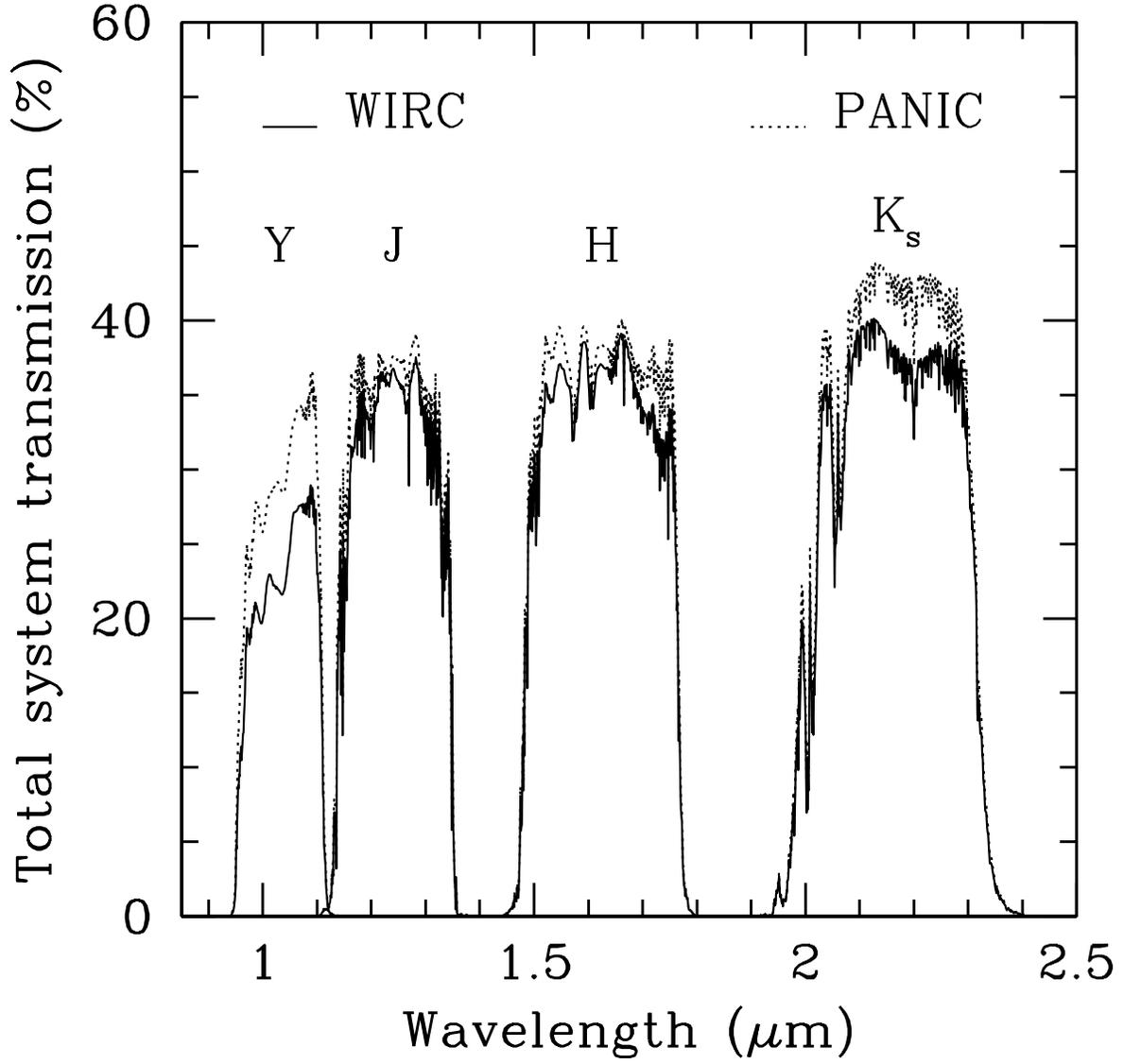}
\caption{Total transmission curves for $YJHK_s$ in WIRC and PANIC. The
curves include the Earth's atmosphere, the telescope and instrument
optical elements, and the detector QE.
\label{IR_filters_fig}}
\end{figure}

\begin{figure}
\plotone{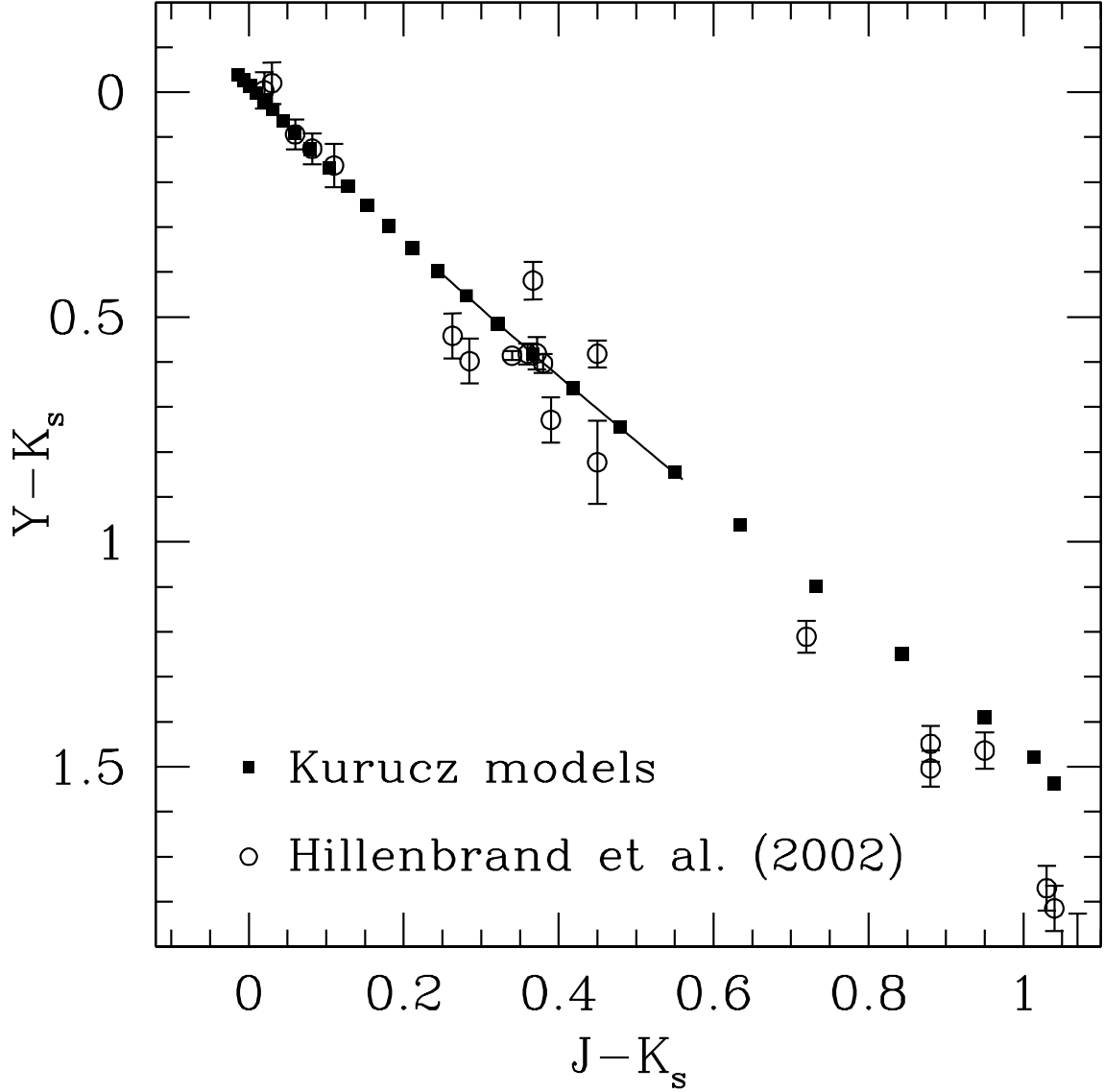}
\caption{($Y-K_s$) vs. ($J-K_s$) diagram for a grid of main-sequence-analog 
Kurucz model atmospheres ({\em filled
squares\,}), and measurements of standard stars published by \citet{hillenbrand02} ({\em open
circles\,}). The solid line shows a polynomial fit to the models
drawn over the range of ($J-K_s$) colors covered by the Persson et~al. (1998) standard stars.}
\label{kurucz_fig}
\end{figure}

In $\S$ \ref{nir_red}, the $Y$-band photometric solution represents a
special case, since \citet{persson98} only defined the standard $JHK_s$
system. The $Y$-band system has been recently introduced by
\citet{hillenbrand02}. In order to derive $Y$ magnitudes for all of the
Persson et al. (1998) standard stars, we defined a relationship between the
($Y-K_s$) and ($J-K_s$) colors for a series of Kurucz model 
spectra\footnote{Main-sequence models with $T_{eff}$ between 3,500 and 10,000 K,
log$~g$ = 4, and solar abundances.}.
These models have solar abundances and effective temperatures between 
3,500 and 10,000 K, thus producing a grid of spectra for main-sequence 
star analogs. We computed synthetic $YJK_s$ magnitudes for each model 
using the total system transmission functions for PANIC shown in 
Figure \ref{IR_filters_fig}. Figure \ref{kurucz_fig} shows
the resulting ($Y-K_s$) vs. ($J-K_s$) diagram together
with the measurements published by \citet{hillenbrand02} (their Table
2). The data from \citet{hillenbrand02} scatter significantly, but in
general they follow the synthetic values for $(J-K_s)<0.5$ mag. For
$0.7<(J-K_s)<1.2$ mag, there is a systematic difference of
$\sim$0.1 mag in ($Y-K_s$) between the measurements and the models.

   We decided to use the synthetic colors from the models to obtain a
fifth-order polynomial fit of the form

\begin{equation}
(Y-K_s) = \sum_{i=0}^5~a_i~(J-K_s)^i,  
\end{equation}

\noindent where $a_0=-0.017$, $a_1=1.901$, $a_2=-1.296$, $a_3=2.289$,
$a_4=-2.409$, and $a_5=0.999$. The fit is shown in
Figure \ref{kurucz_fig} in the range of colors for the
\citet{persson98} standards. We chose to ignore any zero-point
difference and thus defined $(Y-K_s)=0$ when $(J-K_s)=0$ in order to
maintain the original definition of $\alpha$ Lyr having 0.00 magnitude
at all wavelengths \citep{elias82}. We therefore used the following
formula to compute $Y$-band magnitudes from $J$ and $K_s$ for all of
the \citet{persson98} standards: 

\begin{equation}
Y = K_s +  1.901~(J-K_s) - 1.296~(J-K_s)^2 + 2.289~(J-K_s)^3 -
2.409~(J-K_s)^4 + 0.999~(J-K_s)^5.
\end{equation}

As suggested by the referee, we re-computed $Y$-band magnitudes for
the Persson et al. (1998) standards using the $(Y-J)$ versus $(J-H)$
relation yielded by the Kurucz models.  The differences in the $Y$
magnitudes obtained with both methods are never greater than 0.025 mag
and the mean difference is 0.006 $\pm$ 0.007 mag. We conclude that
both approaches ($Y$ from $Y-J$ and $Y$ from $Y-K$) are valid and
yield very similar results.

\clearpage

\clearpage

\clearpage

\clearpage

\clearpage

\clearpage

\clearpage

\clearpage

\clearpage


\clearpage


\clearpage

\clearpage

\clearpage

\clearpage

\begin{deluxetable} {cccccccc}
\tabletypesize{\scriptsize}
\rotate
\tablecolumns{8}
\tablenum{1}
\tablewidth{0pc}
\tablecaption{Instruments used for the low-$z$ CSP\tablenotemark{a} \label{tab_inst}}
\tablehead{
\colhead{Telescope} & 
\colhead{Instrument} & 
\colhead{Detector} & 
\colhead{Plate Scale} &
\colhead{Filters or $\Delta$$\lambda$} &
\colhead{Photometry/} \\
\colhead{} &
\colhead{} &
\colhead{} &
\colhead{(arcsec pixel$^{-1}$)} &
\colhead{(\AA)} &
\colhead{Spectroscopy?} 
} 
\startdata

Swope      &    CCD Camera       &  SITe 2048$\times$3150 (``Site\#3'')    & 0.435 & $u'g'r'i'BV$  & Phot \\
duPont     &    WFCCD            &  Tek  2048$\times$2048 (``TEK\#5'')     & 0.774 & $BVI$         & Phot \\
duPont     &    CCD Camera       &  Tek  2048$\times$2048 (``TEK\#5'')     & 0.259 & $u'g'r'i'BV$  & Phot \\
Clay       &    LDSS-2           &  SITe 2048$\times$2048 (``Site\#1'')    & 0.380 & $BVR$         & Phot \\
Clay       &    LDSS-3           &  STA  4064$\times$4064 (``STAO500A'')   & 0.190 & $Bg'r'i'$     & Phot \\
Swope      &    RetroCam         &  HAWAII-1 1024$\times$1024              & 0.540 & $YJH$         & Phot \\
duPont     &    WIRC             &  HAWAII-1 1024$\times$1024              & 0.196 & $YJHK_s$      & Phot \\
Baade      &    PANIC            &  HAWAII-1 1024$\times$1024              & 0.125 & $YJHK_s$      & Phot \\
duPont     &    WFCCD            &  Tek  2048$\times$2048 (``TEK\#5'')     & 0.774 & 3,800-9,200   & Spec \\
duPont     &Modular Spectrograph &  SITe 1752$\times$572  (``Site\#2'')    & 0.350 & 3,780-7,270   & Spec \\
Clay       &    LDSS-2           &  SITe 2048$\times$2048 (``Site\#1'')    & 0.380 & 3,600-9,000   & Spec \\
CTIO 1.5-m &  R-C Spectrograph   &  LORAL 1200$\times$800                  & 0.270 & 3,000-10,100  & Spec \\

\enddata
\tablenotetext{a}{Additional data are obtained at Lick Observatory with KAIT, the Nickel 1-m 
reflector, and the Shane 3-m telescope; see, e.g., \cite{foley03}.} 
\end{deluxetable}

\clearpage

\begin{deluxetable} {lcc}
\rotate
\tablecolumns{3}
\tablenum{2}
\tablewidth{0pc}
\tablecaption{Specifications for the SDSS and Johnson filters \label{tab_filters}}
\tablehead{
\colhead{Filter} & 
\colhead{Instrument} & 
\colhead{Specifications}}
\startdata

$u'$  &  Swope/CCD Camera   & AR*KG5 (2 mm) + UG11  (1 mm) + Fused silica (3 mm) + Fused silica (4 mm)*IF \\
$g'$  &  Swope/CCD Camera   & AR*GG400 (2 mm) + BG40 (2 mm) + BK7 (6 mm)*IF \\
$r'$  &  Swope/CCD Camera   & AR*OG550 (4 mm) + BK7 (6 mm)*IF  \\
$i'$  &  Swope/CCD Camera   & AR*RG695 (4 mm) + BK7 (6 mm)*IF  \\
$B$   &  Swope/CCD Camera   & GG385 (2 mm) + BG12 (1 mm) +  S8612 (2 mm)                   \\
$V$   &  Swope/CCD Camera   & GG495 (2 mm) + S8612 (3 mm)                                 \\
$B$   &  Clay/LDSS2         & BG12  (1 mm) + BG39 (2 mm) + GG385 (1 mm)                  \\
$V$   &  Clay/LDSS2         & GG495 (2 mm) + BG39 (2 mm)                               \\
$R$   &  Clay/LDSS2         & OG570 (2 mm) + KG3 (2 mm)                                 \\

\enddata
\tablecomments{AR = Anti-reflection coating; IF = Interference film}
\end{deluxetable}

\clearpage

\begin{deluxetable} {lll}
\tablecolumns{3}
\tablenum{3}
\tablewidth{0pc}
\tablecaption{Coefficients for Swope SITe CCD linearity corrections 
\label{tab_coef}}
\tablehead{
\colhead{UT Date} & 
\colhead{$a_2$} & 
\colhead{$a_3$}}
\startdata

2001 May    &   $-$0.1124              & +0.0478            \\
2001 Jul    &   $-$0.1596              & +0.0926            \\
2001 Sep    &   $-$0.1554              & +0.0853            \\
2002 summer &   $-$0.1485$\pm$0.0001   & +0.0823$\pm$0.0004 \\
2003 May    &   $-$0.1573$\pm$0.0006   & +0.0924$\pm$0.0006 \\
2004 May    &   $-$0.1985              & +0.1319            \\
2004 Oct 6  &   $-$0.134$\pm$0.009     & +0.050$\pm$0.009   \\
2004 Nov 10 &   $-$0.138$\pm$0.008     & +0.056$\pm$0.009   \\
2005 Apr 22 &   $-$0.093$\pm$0.013     & +0.012$\pm$0.014   \\
2005 Apr 27 &   $-$0.139$\pm$0.009     & +0.061$\pm$0.009   \\

\enddata
\end{deluxetable}

\end{document}